# On the optimal focusing of solitons and breathers in long wave models

## Alexey Slunyaev

Institute of Applied Physics, Nizhny Novgorod, Russia, Slunyaev@appl.sci-nnov.ru

## Abstract

Conditions of optimal (synchronized) collisions of any number of solitons and breathers are studied within the framework of the Gardner equation with positive cubic nonlinearity, which in the limits of small and large amplitudes tends to other long-wave models, the classic and the modified Korteweg – de Vries equations. The local solution for an isolated soliton or breather within the Gardner equation is obtained. The wave amplitude in the focal point is calculated exactly. It exhibits a linear superposition of partial amplitudes of the solitons and breathers. The crucial role of the choice of proper soliton polarities and breather phases on the cumulative wave amplitude in the focal point is demonstrated. Solitons are most synchronized when they have alternating polarities. The straightforward link to the problem of synchronization of envelope solitons and breathers in the focusing nonlinear Schrödinger equation is discussed (then breathers correspond to envelope solitons propagating above a condensate).

## Keywords

## 1. Introduction

In various realms waves undergo processes of concentration and divergence of energy, though they produce the most effect in focal zones. Focusing of dispersive waves occurs most trivially in the linear setting. Then the classic Fourier transform represents a powerful tool; the Fourier modes preserve energy, and their phases change evenly according to the phase velocities. Nonlinearity complicates the process; the corresponding small corrections to the evolution of Fourier modes still may be taken into account within the weakly nonlinear theory. This approach fails when solitons arise, as then effects of dispersion and nonlinearity are of similar importance. Since solitons have been discovered in plenty of media, see e.g. in [Remoissenet, 1999], collisions of solitons are essential in many tasks.

Focusing of solitons in integrable models may be studied by means of the Inverse Scattering Technique (IST, frequently called the nonlinear Fourier transform, e.g. [Novikov et al, 1984; Drazin & Johnson, 1996; Osborne, 2010]) or related approaches, though this generalization of the Fourier method is much more complicated in use. Even when the field is decomposed using the IST machinery into modes related to invariant spectrum, their evolution is not trivial; for example, soliton structures are known to experience nonlinear phase shifts when interact.

A soliton gas is a brilliant example when collisions between solitons are essential [Zakharov, 1971; El & Kamchatnov, 2005]. One of the achievements of the IST is the proof that the solution depends on all pair combinations of the soliton parameters (specifically, the discrete eigenvalues of the associated scattering problem). In a rarefied soliton gas collisions between solitons occur mostly pairwise, when the existence of other solitons is unimportant, what helps to treat an interacting soliton ensemble as the ensemble of independent pairs of interacting solitons. Then much information about the gas characteristics may be achieved based on the detailed picture of the two-soliton solution [Pelinovsky et al, 2013; Pelinovsky & Shurgalina, 2015; Shurgalina, 2018].

However, collisions of many soliton-like structures can cause such exciting wave phenomenon as so-called super-rogue waves, which occur as a result of the modulational instability and describe amazing local enhancement of waves (they may be modeled by high-order rational breathers of the nonlinear Schrödinger equation [Chabchoub et al, 2012; Slunyaev et al, 2013]). Similar effects may be utilized to amplify the wave field many times, and are routinely used in the practice of generation of intense optical pulses. On the other hand, rogue wave events are unwanted in other applications: when occur in soliton-based transmission systems or in the field of oceanic waves [Kharif et al, 2009; Slunyaev et al, 2011; Onorato et al, 2013]. Hence both the problems of the soliton management demand thorough investigations: how to focus many solitons into a huge spike, and how to avoid this phenomenon.

The optimal synchronization of envelope solitons, i.e., when they superimpose in phase, was considered in [Sun, 2016] in the framework of the focusing nonlinear Schrödinger equation. Collisions of many solitons and breathers within the modified Korteweg – de Vries equation with positive cubic nonlinearity were studied in [Slunyaev & Pelinovsky, 2016], where such processes were suggested as a new mechanism of rogue wave generation. We emphasize two observations made in these works: (i) the magnitude of the field in the focal point is represented by a linear superposition of the partial soliton amplitudes; (ii) a train of long-wave solitons experiences optimal focusing when they have alternating polarities, the sign of the field in the focus is inherited from the fastest soliton. In the present paper we extend the results reported in [Slunyaev & Pelinovsky, 2016] to a more general integrable long-wave model, the focusing Gardner equation, which tends to the classic Korteweg – de Vries equation and to the modified Korteweg – de Vries equation in the limits of small or large waves respectively.

The paper is organized as follows. The Gardner equation and transformations to the other integrable long wave models are presented in Sec. 2. The method of construction of the multisoliton solutions of the Gardner equation is described in Sec. 3; more details on the modification of the Darboux transform are provided in Appendix A. The local expression for an isolated solution is discussed in Sec. 4, while the details of the derivation are given in Appendix B. The field magnitude in the focal point is calculated in Sec. 5 with more details of the derivation moved to Appendix C. In Sec. 6 the general case of collisions between solitons and breathers is discussed. In Conclusions besides the closing remarks we describe the straightforward relation of the obtained results to the problem of focusing of solitons in the nonlinear Schrödinger equation with zero background and when propagating over a condensate.

## 2. The Gardner equation

In this paper we will concern the Gardner equation (GE) with focusing type of nonlinearity in the standard dimensionless form

$$\frac{\partial u}{\partial t} + 6u\frac{\partial u}{\partial x} + 6u^2\frac{\partial u}{\partial x} + \frac{\partial^3 u}{\partial x^3} = 0, \qquad (1)$$

where $u(x,t)$ is a real function of space and time.

Exact reductions of the GE to the Korteweg – de Vries (KdV) and the modified Korteweg – de Vries (mKdV) equations exist. Solutions of the GE $u(x,t)$ are related to solutions $w(x,t)$ of the mKdV equation,

$$\frac{\partial w}{\partial t} + 6w^2 \frac{\partial w}{\partial x} + \frac{\partial^3 w}{\partial x^3} = 0, \qquad (2)$$

via linear transformations of the function and coordinate

$$w(x,t) = \frac{1}{2} + u\left(x - \frac{3}{2}t, t\right), \qquad (3)$$

which adds to the solution a constant pedestal. The complex Miura transformation

$$v(x,t) = u(1+u) + i\frac{\partial u}{\partial x} \qquad (4)$$

links the solutions of the GE and the solutions $v(x,t)$ of the KdV equation

$$\frac{\partial v}{\partial t} + 6v \frac{\partial v}{\partial x} + \frac{\partial^3 v}{\partial x^3} = 0, \qquad (5)$$

which may be complex-valued in the general case due to the imaginary unit in (4).

Meanwhile in the limit of small wave amplitudes equation (1) tends to the KdV equation (5). In the other limit of large waves the quadratic term may be neglected, and equation (1) tends to the focusing mKdV equation (2).

All the listed evolution equations (1), (2), (5) are integrable by means of the Inverse Scattering Technique and possess soliton solutions. The Gardner equation (1) is known to have solitons (see e.g. [Slunyaev, 2001]) in the form

$$u_\pm(x,t) = \frac{4\mu^2}{1 \pm \sqrt{1+4\mu^2} \cosh[2\mu(x - x_0 - 4\mu^2 t)]}, \qquad (6)$$

where $\mu$ is a positive parameter, which corresponds to the spectral data of the associated scattering problem (see in [Grimshaw et al, 2010]), the sign $\pm$ specifies polarity of the soliton, $u_+(x,t) > 0$ and $u_-(x,t) < 0$, and $x_0$ is the position at $t = 0$. The soliton solution (6) is characterized by the amplitude

$$A_\pm = \frac{4\mu^2}{1 \pm \sqrt{1+4\mu^2}} = \pm\sqrt{1+4\mu^2} - 1 \qquad (7)$$

and decays to zero exponentially far from its location ($x_0 + 4\mu^2 t$); the soliton velocity equals to $4\mu^2$. The soliton amplitude grows in absolute value when its velocity increases. Shapes of positive and negative solitons (6) are different. For a given $\mu$ (and hence, for the given velocity), the following relation between the amplitudes of the positive and negative solitons holds,

$$A_+ + A_- = -2. \qquad (8)$$

In the limit $\mu \to 0$ positive solitons (6) tend to the soliton solutions of the KdV equation with the velocity proportional to the amplitude,

$$u_+(x,t) \xrightarrow[\mu \to 0]{} \frac{2\mu^2}{\cosh^2[\mu(x - x_0 - 4\mu^2 t)]}. \qquad (9)$$

The competition between two nonlinear terms in (1) leads to existence of an algebraic negative soliton when $\mu \to 0$ with the amplitude $A_- = -2$. The heights of negative solitons are noticeably higher than the heights of positive solitons with the same velocity; negative solitons cannot be smaller in heights than the algebraic soliton.

In the limit of large-amplitude waves the soliton solution (6) yields

$$u_\pm(x,t) \xrightarrow[|\mu| \gg 1]{} \pm \frac{2\mu}{\cosh[2\mu(x - x_0 - 4\mu^2 t)]}, \tag{10}$$

which is the soliton solution of the mKdV equation. These solitons may have either polarity and are symmetric. The velocity is proportional to the squared soliton amplitude.

## 3. Construction of multisoliton solutions of the Gardner equation with the help of the Darboux transform

The Darboux transform is a powerful method for construction of multisoliton solutions, see e.g. the book [Matveev & Salle, 1991]. The Darboux transform for the KdV equation was adapted in [Kulikov & Fraiman, 1996] to the modified KdV equation. This approach was developed further in [Slunyaev & Pelinovsky, 1999] and [Slunyaev, 2001] bearing in mind the simple relation (3) between the mKdV and the Gardner equations. Two-soliton solutions were derived there; the 3-soliton solution for the Gardner equation (1) was obtained in [Chow et al, 2005] with the use of the Hirota method.

The function $u(x,t)$ is a solution of the Gardner equation (1) when it is obtained through the ratio of two Wronskians,

$$u(x,t) = i\partial_x \ln \frac{W_{N+1}}{W_N} - \frac{1}{2}, \tag{11}$$

$$W_N(\psi_1, \psi_2, ..., \psi_N) = \begin{vmatrix} \psi_1 & \psi_2 & \cdots & \psi_N \\ \partial_x \psi_1 & \partial_x \psi_2 & \cdots & \partial_x \psi_N \\ \vdots & \vdots & \ddots & \vdots \\ \partial_x^{N-1} \psi_1 & \partial_x^{N-1} \psi_2 & \cdots & \partial_x^{N-1} \psi_N \end{vmatrix},$$

$$W_{N+1}(\psi_0, \psi_1, \psi_2, ..., \psi_N) = \begin{vmatrix} \psi_0 & \psi_1 & \cdots & \psi_N \\ \partial_x \psi_0 & \partial_x \psi_1 & \cdots & \partial_x \psi_N \\ \vdots & \vdots & \ddots & \vdots \\ \partial_x^N \psi_0 & \partial_x^N \psi_1 & \cdots & \partial_x^N \psi_N \end{vmatrix}.$$

The so-called 'seed' functions $\psi_j$, $j = 1, ..., N$, are either *cosh* or *sinh* functions,

$$\psi_j(x,t) = \cosh[\Theta_j + i\theta_j] \quad \text{or} \quad \psi_j(x,t) = \sinh[\Theta_j + i\theta_j], \quad \Theta_j \equiv \mu_j(x - \mu_j^2 t). \tag{12}$$

We assume for certainty that $\text{Re}(\mu_j) > 0$. The function $\psi_0$ is defined as

$$\psi_0 = \exp\left(-\frac{i}{2}(x+t)\right). \tag{13}$$

In (11) $\partial_x$ denotes differentiation with respect to $x$. In the general case the functions (12) and (13) could have arbitrary shifts of the coordinate similar to $x_0$ in (6), but we assume all of them to be zeros. This assumption specifies locations of the solitons at $t = 0$ as will be seen in Sec. 4. Note that relabeling of $\psi_j$ does not influence the eventual solution (11). The procedure of application of the Darboux transform to the Gardner equation is described in more detail in Appendix A.

Note that the seed functions (12) may be represented in an alternative formalized form through an imaginary phase shift,

$$\psi_j(x,t) = i^{l_j} \cosh\left[\Theta_j + i\theta_j - i\frac{\pi}{2}l_j\right], \tag{14}$$

where $l_j$ are integer numbers. Even $l_j$ correspond to the choice *cosh* in (12), while odd $l_j$ – to *sinh*. Note that any constant coefficients before the functions $\psi_j$ in (11) do not change the solution $u(x,t)$, therefore the multiplier $i^{l_j}$ could be removed in (14).

The values $\theta_j$ of the additional complex phase shift in (12), (14) are defined by the relations

$$\theta_j = \frac{1}{2}\arctan 2\mu_j. \qquad (15)$$

Thus, $-\pi/2 \leq \theta_j \leq \pi/2$ if $\mu_j$ is real. The choice (15) will be grounded later in Sec. 4.

For our purposes it is more convenient to rewrite solution (11) in the form

$$u(x,t) = i\frac{\partial_x W_{N+1}}{W_{N+1}} - i\frac{\partial_x W_N}{W_N} - \frac{1}{2} \qquad (16)$$

and to use the formalized definition of the determinant operation. The determinant of matrix $D$ of the size $N \times N$ with elements $d_{ij}$ may be calculated as follows,

$$\det D = \sum_{(\alpha_1, \alpha_2, \ldots, \alpha_N)} (-1)^{p_\alpha} d_{\alpha_1 1} d_{\alpha_2 2} \cdot \ldots \cdot d_{\alpha_N N}. \qquad (17)$$

Here the summation is performed along all possible permutations ($\alpha_1$, $\alpha_2$, ..., $\alpha_N$) of the sequence of natural numbers (1, 2, ..., $N$) (in total $N!$ permutations, hence (17) consists of $N!$ summands), and $p_\alpha$ is the number of inversions in the permutation from (1, 2, ..., $N$) to ($\alpha_1$, $\alpha_2$, ..., $\alpha_N$). With the use of this representation the Wronskian in (11) may be written in the following form,

$$W_N = \sum_\alpha c_\alpha \psi_{\alpha_1} \partial_x \psi_{\alpha_2} \partial_x^2 \psi_{\alpha_3} \cdot \ldots \cdot \partial_x^{N-1} \psi_{\alpha_N} = \sum_\alpha c_\alpha \mathrm{M}_\alpha \Psi_\alpha(x,t), \qquad (18)$$

where new notations have been introduced for brevity,

$$c_\alpha \equiv (-1)^{p_\alpha}, \qquad \mathrm{M}_\alpha \equiv \mu_{\alpha_2} \mu_{\alpha_3}^2 \mu_{\alpha_4}^3 \cdot \ldots \cdot \mu_{\alpha_N}^{N-1},$$

$$\Psi_\alpha(x,t) \equiv \psi_{\alpha_1} \overline{\psi}_{\alpha_2} \psi_{\alpha_3} \overline{\psi}_{\alpha_4} \cdot \ldots \cdot \begin{cases} \psi_{\alpha_N}, & N \text{ is odd} \\ \overline{\psi}_{\alpha_N}, & N \text{ is even} \end{cases}. \qquad (19)$$

Note that $c_\alpha$ and $\mathrm{M}_\alpha$ do not depend on $x$ or $t$. In (19) the overlines by definition mean the following,

$$\overline{\psi}_j \equiv \frac{1}{\mu_j}\frac{\partial \psi_j}{\partial x}, \qquad j = 0, 1, \ldots, N. \qquad (20)$$

In terms of (14) the overline corresponds to the increase of $l_j$ by 1. Note that (20) is applicable to $\psi_0$ as well, where we define for generality $\mu_0 = -i/2$. Then $\overline{\psi}_0 = \psi_0$.

Similar to (18)-(19), we define

$$W_{N+1} = \sum_\beta c_\beta \mathrm{M}_\beta \Psi_\beta(x,t) \qquad (21)$$

$$c_\beta \equiv (-1)^{p_\beta}, \qquad \mathrm{M}_\beta \equiv \mu_{\beta_1} \mu_{\beta_2}^2 \mu_{\beta_3}^3 \cdot \ldots \cdot \mu_{\beta_N}^N,$$

$$\Psi_\beta(x,t) \equiv \psi_{\beta_0} \overline{\psi}_{\beta_1} \psi_{\beta_2} \overline{\psi}_{\beta_3} \cdot \ldots \cdot \begin{cases} \overline{\psi}_{\beta_N}, & N \text{ is odd} \\ \psi_{\beta_N}, & N \text{ is even} \end{cases}. \qquad (22)$$

The summation in (21) is performed along all possible permutations ($\beta_0$, $\beta_1$, ..., $\beta_N$) of the sequence of natural numbers (0, 1, 2, ..., $N$), and $p_\beta$ is the number of inversions in the permutation from (0, 1, 2, ..., $N$) to the sequence ($\beta_0$, $\beta_1$, ..., $\beta_N$). The quantities $c_\beta$ and $\mathrm{M}_\beta$ are constants.

## 4. Local solution for an isolated soliton

In this section we assume that all $\mu_j$, $j = 1, \ldots, N$, are different real numbers. Let us consider the $N$-soliton solution specified by the Darboux transform (11)-(13), supposing that at some time the soliton $n$ is located far from all the other solitons having $L$ solitons on the left and $R$ solutions on the right, $L + R + 1 = N$. Mathematically this means that in the considered moments of time

$$\Theta_j \ll \Theta_n \quad \text{for} \quad 1 \leq j \leq L; \quad \text{and} \quad \Theta_n \ll \Theta_j \quad \text{for} \quad r \leq j \leq N, \quad r \equiv N - R + 1. \tag{23}$$

We numerate the solitons in such a way that solitons with numbers from 1 to $n$ are located on the left, and from $n+1$ to $N$ – on the right. Then one may apply such change of references that $\Theta_n = O(1)$, and the following asymptotics is valid for $j \neq n$,

$$\overline{\psi}_j \approx \psi_j \approx \exp[\Theta_j + i\theta_j], \quad \text{for} \quad 1 \leq j \leq L,$$
$$-\overline{\psi}_j \approx \psi_j \approx \exp[-\Theta_j - i\theta_j], \quad \text{for} \quad r \leq j \leq N. \tag{24}$$

The details of the following calculations are given in Appendix B. They result in the local solution in the form (see (B.15))

$$u = i\mu_n \cdot \begin{cases} \tanh\left(\Theta_n + i\left(\theta_n + \dfrac{\varphi_n}{2}\right) + \dfrac{1}{2}\ln\chi_n\right) - \tanh\left(\Theta_n + i\theta_n + \dfrac{1}{2}\ln\chi_n\right), & \text{if} \quad \chi_n > 0 \\ \coth\left(\Theta_n + i\left(\theta_n + \dfrac{\varphi_n}{2}\right) + \dfrac{1}{2}\ln|\chi_n|\right) - \coth\left(\Theta_n + i\theta_n + \dfrac{1}{2}\ln|\chi_n|\right), & \text{if} \quad \chi_n < 0 \end{cases}, \tag{25}$$

when the seed function is $\psi_n = \cosh(\Theta_n + i\theta_n)$. If $\psi_n = \sinh(\Theta_n + i\theta_n)$, then in (25) functions *tanh* and *coth* should be swapped. The parameter $\chi_n$

$$\chi_n = \frac{\prod_{j=1}^{L}(\mu_j - \mu_n) \prod_{j=N-R+1}^{N}(\mu_j + \mu_n)}{\prod_{j=1}^{L}(\mu_j + \mu_n) \prod_{j=N-R+1}^{N}(\mu_j - \mu_n)}, \qquad \chi_1 = 1 \quad \text{if} \quad N = 1 \tag{26}$$

controls the shifts of coordinates, which depend on parameters of the solitons and on their mutual location. The sign of $\chi_n$

$$s_n \equiv \operatorname{sgn} \chi_n = \operatorname{sgn} \prod_{j=1, j \neq n}^{N}(\mu_j^2 - \mu_n^2). \tag{27}$$

depends on the pairs of $\mu_j$, though is not affected by locations of the solitons.

The request on the solution $u(x,t)$ to be real-valued for arbitrary $\mu_n$, imposes particular relations between $\varphi_n$ and $\theta_n$, $\theta_n + \varphi_n/2 = -\theta_n$, what is fulfilled if $\theta_j$ are defined according to (15). Then the solution (25) eventually tends to the following,

$$u(x,t) = \begin{cases} -i\mu_n \dfrac{\overline{\psi}_n\left(\Theta_n + i\theta_n + \dfrac{1}{2\mu_n}\ln\chi_n\right)}{\psi_n\left(\Theta_n + i\theta_n + \dfrac{1}{2\mu_n}\ln\chi_n\right)} + c.c., & \text{if} \quad \chi_n > 0 \\ -i\mu_n \dfrac{\psi_n\left(\Theta_n + i\theta_n + \dfrac{1}{2\mu_n}\ln|\chi_n|\right)}{\overline{\psi}_n\left(\Theta_n + i\theta_n + \dfrac{1}{2\mu_n}\ln|\chi_n|\right)} + c.c., & \text{if} \quad \chi_n < 0 \end{cases}. \tag{28}$$

The solution may be represented in a shorter form with the use of (14), see (B.18). Note that the expressions (28) coincide with the solution (6) for a single soliton except for the shifts of coordinate, when the following equalities are employed,

$$\tanh i\theta_n = \frac{2i\mu_n}{1 + \sqrt{1 + 4\mu_n^2}}, \quad \coth i\theta_n = \frac{2i\mu_n}{1 - \sqrt{1 + 4\mu_n^2}}. \tag{29}$$

The calculated shifts of the soliton locations coincide with the ones derived for the mKdV equation [Wadati, 1973] and the Gardner equation with negative cubic nonlinearity [Romanova, 1979]. With the use of (29) the expressions for soliton amplitudes (7) may be represented in the alternative convenient form

$$A_+ = 2\mu \tan\theta, \quad A_- = 2\mu \cot\theta, \quad \theta = \frac{1}{2}\arctan 2\mu. \tag{30}$$

To conclude, the following solution is valid in the vicinity of the $n$-th soliton:

$$u(x,t) = u_+\left(x + \frac{1}{2\mu_n}\ln|\chi_n|, t\right) \quad \text{if } \psi_n = \cosh[\Theta_n + i\theta_n] \text{ and } \chi_n > 0$$

$$\text{or if } \psi_n = \sinh[\Theta_n + i\theta_n] \text{ and } \chi_n < 0;$$

$$u(x,t) = u_-\left(x + \frac{1}{2\mu_n}\ln|\chi_n|, t\right) \quad \text{if } \psi_n = \cosh[\Theta_n + i\theta_n] \text{ and } \chi_n < 0$$

$$\text{or if } \psi_n = \sinh[\Theta_n + i\theta_n] \text{ and } \chi_n > 0. \tag{31}$$

In the case of a single soliton, $N = 1$, $\chi_1 = 1$, and hence the coordinate shift vanishes in (31).

As at $t \to \pm\infty$ the solitons always get separated due to the difference in their velocities, the solution at these limits is described by (31), and thus real valued, though the expression for the multisoliton solution (11) contains the imaginary unit. A real-valued Cauchy problem for the GE remains real, hence the obtained multisoliton solution is real at any time.

If in the limit $t \to -\infty$ the $n$-th soliton has $L$ solitons from the left and $R$ solitons from the right, it is obvious that when $t \to +\infty$, it will have $R$ solitons from the left and $L$ solitons from the right. Thus from (26) we have $\ln|\chi_n(t \to +\infty)| = -\ln|\chi_n(t \to -\infty)|$ for any $n$-th soliton. Therefore the symmetry $u(-x,-t) = u(x,t)$ exists in the limit of large times. Hence, the condition $x = 0$, $t = 0$ corresponds to an especial point of the solution (11), which describes a collision of many solitons. Hereafter we will refer to this point as the focal point.

Due to the revealed symmetry (which is caused by the chosen zero coordinate shifts in the seed functions (12)), the considered solution (11)-(13) is obviously not the general $N$-soltion solution if $N > 2$. The solution represents the general $N$-soliton solution when the arbitrary global shift of coordinate $x' = x - x_0$ is allowed only if $N = 1$ or $N = 2$.

## 5. Focal point of the multisoliton solution

Let us calculate the value $u(x=0, t=0)$ from (16), assuming that $\mu_j$, $j = 1, \ldots, N$, are different real numbers. The details of the calculations are given in Appendix C. The final result may be written in the form (C.10)

$$u = \begin{cases} -2i\mu_1 \dfrac{\overline{\psi_1}}{\psi_1} + 2i\mu_2 \dfrac{\psi_2}{\overline{\psi_2}} - 2i\mu_3 \dfrac{\overline{\psi_3}}{\psi_3} + \ldots - 2i\mu_N \dfrac{\overline{\psi_N}}{\psi_N}, & N \text{ is odd} \\ -2i\mu_1 \dfrac{\psi_1}{\overline{\psi_1}} + 2i\mu_2 \dfrac{\overline{\psi_2}}{\psi_2} - 2i\mu_3 \dfrac{\psi_3}{\overline{\psi_3}} + \ldots + 2i\mu_N \dfrac{\overline{\psi_N}}{\psi_N} & N \text{ is even} \end{cases} \quad \text{at } (x=0, t=0). \tag{32}$$

$N$ solitons interact due to the difference between their velocities, proportional to $\mu_j^2$. Long before the collisions the $n$-th soliton is specified by the relation (31), and its sign depends on the relations between $\mu_j^2$ and $\mu_n^2$ through the parameter $\chi_n$ (26).

As has been mentioned before, a change of the order of seed functions $\psi_j$ does not alter the solution $u(x,t)$. In what follows let us numerate the solitons in the order of descending velocities, when they are arranged at $t \to -\infty$, i.e., $\mu_j > \mu_{j+1}$ for $j = 1, \ldots, N-1$. The soliton $n = 1$ is then the fastest, and for it $\text{sgn}\chi_1 = (-1)^{N-1}$ – hence for the given seed function its polarity depends on the total number of solitons constructed with the help of (11).

If only one soliton exists, $N = 1$, and $\chi_1 = 1 > 0$, then (32) yields

$$u(0,0) = -2i\mu_1 \frac{\overline{\psi}_1}{\psi_1}. \tag{33}$$

Assume that the seed function for $\psi_1$ is *cosh*. Then according to (30) and (31) the soliton is positive with amplitude $A_{1+} = 2\mu_1 \tan\theta_1$. Meanwhile from (33) $u(0,0) = -2i\mu_1 \tanh i\theta_1 = 2\mu_1 \tanh \theta_1$. Thus, $u(0,0) = A_{1+}$ for the chosen seed function. If the seed function if *sinh*, then similar calculations result in $u(0,0) = A_{1-}$. Therefore in the general case we obtain a trivial formula

$$u(0,0) = A_1, \text{ when } N = 1, \tag{34}$$

where $A_1$ is the amplitude of the soliton (either positive or negative).

If $N = 2$, then $\chi_1 < 0$ and $\chi_2 > 0$. If the seed function for $\psi_1$ is *cosh*, then due to (30), (31) the first soliton is negative with amplitude $A_{1-} = 2\mu_1 \cot\theta_1 < 0$. Meanwhile the first summand in (32) for even $N$ gives $-2i\mu_1 \psi_1(0,0)/\overline{\psi}_1(0,0) = -2i\mu_1 \coth i\theta_1 = 2\mu_1 \cot \theta_1 = A_{1-}$. Following this way, one may obtain that (32) results in the following solution for the focusing point,

$$u(0,0) = A_1 - A_2, \text{ when } N = 2, \tag{35}$$

where $A_j$, $j = 1, 2$ are the amplitudes of the solitons, which may be positive or negative. This relation was obtained for the two-soliton solution of the GE in [Slunyaev, 2001].

In the general case of $N$ interacting solitons with amplitudes $A_j$, $j = 1, .., N$, which are arranged in the descending order of velocities, the solution (32) yields a simple formula

$$u(0,0) = A_1 - A_2 + A_3 - A_4 + ... + (-1)^{N-1} A_N. \tag{36}$$

Since solitons of either polarity grow in height as the corresponding parameter $\mu$ grows (see (7)), if all the solitons have the same polarity then $|u(0,0)| \leq |A_1|$. Hence when solitons have the same polarity, the field in the focal point does not exceed in magnitude the height of the fastest soliton. In particular, this statement suggests that collisions of the KdV solitons (as the small-amplitude limit of the solution of the GE) can nether produce in the focal point a wave with larger amplitude.

Collisions of solitons with same polarities are shown in Fig. 1 and Fig. 2, where the snapshots of $u(x,t)$ are given long before and right in the moment when the focusing should occur. The general views of the field evolution are shown in Fig. 1b, 2b. Though the waves in Fig. 1a and Fig. 2a look noticeably different due to the asymmetry of positive and negative solitons, the traces of the solitons are same as the shifts of location do not depend on the soliton polarity (cf. Fig. 1b and Fig. 2b). It is seen that in both the cases the modulus of the maximum field amplitude decreases during the interaction (Fig. 1c and Fig. 2c).

If the arranged in the descending order of velocities solitons have alternating polarities, the value of $u(0,0)$ is equal to the linear superposition of the solitons' heights and thus may much exceed the field before the focusing. It is easy to see from (36) that in the situation of alternating soliton polarities the sign of $u(0,0)$, is inherited from the first (i.e., fastest) soliton. Such cases are shown in Fig. 3 and Fig. 4. In the figures the sequences of solitons are characterized by the same velocities as in Fig. 1, 2, but the polarities are alternating having the leftmost (fastest) soliton positive in Fig. 3 and negative in Fig. 4. At the moment $t = 0$ the solitons focus in huge waves with amplitudes much exceeding the maximum amplitude of the solitons in concord with (36). Figs. 3, 4 confirm that the condition $x = 0$, $t = 0$ indeed corresponds to the focal point. In Fig. 3a the maximum wave is positive, while all the positive solitons have heights much smaller than it. Note the qualitative difference between the traces of solitons in Figs. 1b, 2b and Figs. 3b, 4b. The integrals of motion of the GE inevitably require the focused wave to be much compressed in length. The effect of collision of two opposite solitons which results in the wave amplification with change of the sign of the maximum wave was pointed out in [Slunyaev, 2001]. Such 'absorb-emit' collisions were also considered in [Shurgalina & Pelinovsky, 2016] as the reason for the generation of abnormally high waves in a soliton gas.

The presented scenarios of soliton collisions are similar to the ones depicted in [Slunyaev & Pelinovsky, 2016] in the framework of the modified KdV equation, which is a large amplitude limit of the Gardner equation. In particular, the relation similar to (36) was presented in [Slunyaev & Pelinovsky, 2016] without details of the calculation. The proof of this relation for the *N*-soliton solutions of the mKdV equation (A.10) may be performed directly in the spirit of calculations listed in Appendices B and C; though it is significantly more straightforward.

## 6. Collisions of breathers and solitons

It was shown in [Slunyaev, 2001] that the choice of the soliton parameters in the two-soliton solution (11)

$$\mu_1 = \frac{a+ib}{2}, \quad \mu_2 = \frac{a-ib}{2}, \quad a>0, \quad b>0 \tag{37}$$

results in construction of a breather solution of the Gardner equation, which may be presented in the compact form

$$u_{br}(x,t) = \frac{-4\,\mathrm{Im}(\mu^2)}{\mu\,\tanh(\mu(x-4\mu^2 t)+\delta) - \mu^*\,\tanh(\mu^*(x-4\mu^{*2} t)-\delta^*)}, \quad \delta \equiv \frac{1}{4}\ln\frac{i/2+\mu}{i/2-\mu}. \tag{38}$$

The expression (38) is real-valued; it represents the general breather solution if arbitrary changes of the reference coordinate and time are taken into account. The dependence of the solution (38) is governed by the *tanh* functions; the kinematic properties of breathers may be obtained straightforwardly analyzing the phase $\Theta = \mu(x - 4\mu^2 t)$. Its real part controls the envelope, $\mathrm{Re}\,\Theta \propto x - V_{br}t$, while the imaginary part describes the oscillations, $\mathrm{Im}\,\Theta \propto x - \omega_{br}t$. Here $V_{br}$ and $\omega_{br}$ are the velocity and frequency of the breather,

$$V_{br} = a^2 - 3b^2, \qquad \omega_{br} = 3a^2 - b^2. \tag{39}$$

Depending on the parameters, in the chosen reference a breather can propagate rightwards similar to all solitons or leftwards as quasilinear dispersive waves. In the first case a breather looks like two bound solitons of opposite polarities which collide periodically. In the latter case the breather solution resembles an oscillatory wave packet; its amplitude decreases when the imaginary part of $\mu$ grows (see in [Pelinovsky & Grimshaw, 1997; Slunyaev, 2001]). The time when the breather repeats its shape, $T_{br}$, is defined by the full repetition of the phase $\Theta$ taking into account the drift of the envelope with velocity $V_{br}$. This condition reads $T_{br}(V_{br} - \omega_{br})\,b/2 = 2\pi$, which yields

$$T_{br} = \frac{2\pi}{b(a^2+b^2)}. \tag{40}$$

Let us now consider the general form of the solution (11) for given *N* parameters $\mu_j$, $j = 1, \ldots, N$ when a pair $\mu_a$ and $\mu_b$ are complex conjugated, $\mu_b = \mu_a^*$. Then, similar to as done in Sec. 4, we consider the vicinity of such coordinates and times that

$$\Theta_j \ll \mathrm{Re}(\Theta_a), \quad \Theta_j \ll \mathrm{Re}(\Theta_b) \quad \text{for} \quad 1 \leq j \leq L,$$
$$\mathrm{Re}(\Theta_a) \ll \Theta_j, \quad \mathrm{Re}(\Theta_b) \ll \Theta_j \quad \text{for} \quad N-R+1 \leq j \leq N. \tag{41}$$

Here we have *L* solitons on the left and *R* solitons on the right, $R + L + 2 = N$. Then the asymptotics (24) may be applied, and the derivation given in Appendix B may be repeated for the local solution specified by the seed functions $\psi_a$ and $\psi_b$, which leads to the real-valued solution

$$u = i(\mu^2 - \mu^{*2}) \left[ \left( \mu \frac{\overline{\psi}_a\left(\Theta - i\theta + \frac{1}{2}\ln\chi\right)}{\psi_a\left(\Theta - i\theta + \frac{1}{2}\ln\chi\right)} - \mu^* \frac{\overline{\psi}_b\left(\Theta^* - i\theta^* + \frac{1}{2}\ln\chi^*\right)}{\psi_b\left(\Theta^* - i\theta^* + \frac{1}{2}\ln\chi^*\right)} \right)^{-1} - \right.$$

$$\left. - \left( \mu \frac{\overline{\psi}_a\left(\Theta + i\theta + \frac{1}{2}\ln\chi\right)}{\psi_a\left(\Theta + i\theta + \frac{1}{2}\ln\chi\right)} - \mu^* \frac{\overline{\psi}_b\left(\Theta^* + i\theta^* + \frac{1}{2}\ln\chi^*\right)}{\psi_b\left(\Theta^* + i\theta^* + \frac{1}{2}\ln\chi^*\right)} \right)^{-1} \right], \quad (42)$$

$$\Theta \equiv \mu(x - \mu^2 t), \quad \theta = \frac{1}{2}\arctan 2\mu, \quad \chi = \frac{\prod_{j=1}^{L}(\mu_j - \mu) \prod_{j=N-R+1}^{N}(\mu_j + \mu)}{\prod_{j=1}^{L}(\mu_j + \mu) \prod_{j=N-R+1}^{N}(\mu_j - \mu)}, \quad (43)$$

where the low index "$a$" is omitted for brevity, $\mu = \mu_a$, $\Theta = \Theta_a$, $\theta = \theta_a$, $\chi = \chi_a$. One may see that relations (43) agree with (15) and (26), though now numbers $\theta$ and $\chi$ are complex. Besides, the solution (43) is valid not only for real $\mu_j$, but also for complex conjugated pairs of $\mu_j$ with indices in the intervals $1 \leq j \leq L$ or $N-R+1 \leq j \leq N$.

To ensure real-valued solutions (42), $\psi_a$ and $\psi_b$ given by (12) should have the same functional dependence (i.e., *cosh* or *sinh* both). If one chooses $\psi_a$ and $\psi_b$ described by *cosh*, the expression (42) after trigonometric transformations tends to the solution $u_{br}$ (38) with the extra addition $1/2 \ln\chi$ to the argument of *tanh*, inherited from (42). If $\chi = 1$, then (42) and (38) are identical. If, instead, the functions $\psi_a$ and $\psi_b$ are both *sinh*, then in the breather solution (38) functions *tanh* should be replaced by *coth*. Note that $\chi$ is complex-valued, thus besides the shift of location (controlled by the real part of $\Theta$), interactions with other solitons or breathers result in the shift of the breather phase described by the imaginary part of $\Theta$. In particular, as $\ln(-1) = i\pi$, the change of sign of $\chi$ to the opposite effectively changes in the solution *tanh* to *coth* or inversely.

Note that the derivation of (32) remains valid for complex parameters $\mu_j$. If a breather is alone, $N = 2$, $\mu_1 = \mu_a = \mu$, $\mu_2 = \mu_b = \mu^*$, then the amplitude of the breather comes from (32),

$$B \equiv u_{br}(0,0) = -2i\mu \frac{\psi(i\theta)}{\overline{\psi}(i\theta)} + 2i\mu^* \frac{\overline{\psi}(i\theta^*)}{\psi(i\theta^*)}, \quad (44)$$

where the parameter $B$ stands for the breather magnitude in the focal point. With the use of (29) the expression (44) tends to the form

$$B = \mp 2 \operatorname{Re}\sqrt{1 + 4\mu^2}. \quad (45)$$

In (45) $B$ is negative if $\psi_a = \cosh(\Theta + i\theta)$ and is positive in the other case $\psi_a = \sinh(\Theta + i\theta)$. Hence, the quantity $|B|$ does not depend on the choice of the seed function and corresponds to the amplitude of the breather. Note that (45) may be written in the form $B = \mp(A_+ - A_-)$, where $A_+$ and $A_-$ are the formal amplitudes of the positive and negative solitons with the complex parameter $\mu$ obtained according to (7). The breather amplitude $|B|$ is smaller than the difference between the heights of solitons of different polarities with parameters $\operatorname{Re}\mu$. Thus, non-zero values of the imaginary part of $\mu$ result in the decrease of amplitude of the coherent group.

As the expression (32) is valid for complex $\mu_j$, it may be used in the general case when solitons and breathers propagate simultaneously (the simplest case of interaction between a soliton and a breather was studies in [Chow et al, 2005]). The amplitudes of solitons are given by (33) and (34), while the amplitudes of breathers are defined by (44). Then the solution in the focal point (36) may be generalized for the case of $Q$ solitons and $M$ breathers as follows,

$$u(0,0) = A_1 - A_2 + A_3 - A_4 + \ldots + (-1)^{Q-1} A_Q + (-1)^Q (B_1 + B_2 + \ldots + B_M), \qquad (46)$$

where solitons are numerated in the descending order of their velocities.

The formula (46) reveals a difference between the processes of synchronization between solitons and between breathers. The sequence of focusing solitons should be characterized by alternating signs to cause a larger wave, though breathers superimpose in the focal point if they have parameters $B$ of the same sign. We note that an addition of one soliton (or any odd number) to the solution (11) effectively changes the sign of the focused breathers due to the factor $(-1)^Q$ in (46). At the same time breathers retain the dynamics of focusing solitons qualitatively unchanged, and contribute to the focused wave amplitude with their partial amplitudes in the linear manner.

Collisions between two breathers are illustrated in Fig. 5. Parameters of the breathers are chosen in such a way that they have similar amplitudes: $|B_1| \approx 2.8$ for the breather on the left and $|B_2| \approx 2.35$ for the other breather. The breather on the right has effectively larger imaginary part of $\mu$, which results in negative velocity $V_{br}$, while the breather on the left drifts rightwards. The breathers in Fig. 5a-c are characterized by different phases, which in the absence of the other breather result in positive ($B > 0$) or negative ($B < 0$) values of the field in the focal point as indicated in the caption. Depending on the phase combinations the breathers focus at $x = 0$ in large waves of positive (when both $B_1$ and $B_2$ are positive, Fig. 5a) or negative sign (when both $B_1$ and $B_2$ are negative, Fig. 5b). When the signs on $B_1$ and $B_2$ are different, the breathers' amplitudes are subtracted (Fig. 5c).

Fig. 6a and Fig. 6b display the solutions (11) which contain the same breathers as in Fig. 5a and Fig. 5b respectively, and one soliton with small velocity. The soliton has positive (Fig. 6a, the location is shown with the arrow) or negative (Fig. 6b) polarity. Note that comparing the curves for $t = -50$ in Fig. 5a and Fig. 6a (and the same is valid for Fig. 5b and Fig. 6b), the breathers have slightly different locations and noticeably different phases. It is obvious that if the soliton and the other breather are artificially cut out at $t = -50$, then the remaining breather in the focal point will possess the shift $\ln\chi$ according to the solution (42), not (44). Therefore the parameter $B$ is actually not as instructive for the description of focusing breathers. Note that the main effect is produced seemingly not by the soliton amplitude, which is quite small in Fig. 6a, but by the nonlinear shift.

Fig. 7 and Fig. 8 demonstrate the optimal focusing of two breathers and two solitons, which result in huge positive and negative waves respectively. Note that the parameters $\mu$ of the corresponding solitons and breathers in Fig. 7 and Fig. 8 coincide (and thus the velocities are the same as well). In Fig. 7 and Fig. 8 the breathers have opposite phases and the solitons have opposite polarities, hence the amounts $|A_1 - A_2|$, which characterize the heights of the focused waves, are the same in the figures.

Various combinations of focusing solitons and breathers may be suggested by (11) to perform the optimal focusing, if the parameters are chosen in compliance with the described features. Several breathers may have the same velocities and hence can form complicated nonlinear bound patterns; they can also be coupled with solitons of the same celerity. Formula (46) represents the general key rule how to select the signs of particular solitons and phases of breathers to ensure that the interference in the focal point is optimal.

## 7. Conclusions

The processes of collision of many solitons and breathers are considered within the framework of the long-wave Gardner equation with the focusing type of nonlinearity. The case of the KdV equation may be considered as the small-amplitude limit of the studied problem. The height of negative solitons of the Gardner equation is limited from below, and hence in the small-amplitude limit the branch of negative solitons vanishes. Breather solutions of the Gardner equation can possess small amplitudes for sufficiently large imaginary parts of the spectral parameter $\mu$, though such localized solutions do not exist within the framework of the KdV equation. This seeming contradiction is explained by the fact that these coherent structures are supported by the balance between the cubic nonlinearity and dispersion, and hence the corresponding nonlinear term in the equation cannot be neglected. In the large-amplitude limit the Gardner model tends to the mKdV equation.

We consider a particular form of the multisoliton/multibreather solution imposing the condition of time and space symmetry, $u(x,t) = u(-x, -t)$, which helps to synchronize many solitons in one location. The central point of the symmetry, $x = 0$, $t = 0$, is referred to as the focal point and the solution in this point, $u(0,0)$, is calculated exactly in the general case (46). We show that in the focal point the amplitudes of partial solitons and breathers superimpose linearly with the signs provided by the formula (46). This is the main result of the paper. We call the focusing optimal when all the solitons and the breathers are properly synchronized and the wave amplitude in the focal point is maximal. We show that the result of the focusing depends crucially on polarities of the solitons and phases of the breathers. The formula (46) gives the clues how to prepare the conditions for the optimal focusing. In particular, the solitons which tend to focus, besides proper locations accounting for the velocities and nonlinear shifts, should have alternating polarities; the sign of the solution $u(0,0)$ is the same as the sign of the fastest soliton. The wave amplitude in the focal point does not exceed the height of the fastest soliton if focusing solitons are characterized by the same polarities; in this case the fastest soliton is also the largest in amplitude. It is significant that here we discuss the wave amplitude in the focal point only; the global wave maximum may exceed this value, what happens, for example, in the situations of nonoptimal focusing (shown in Fig. 1,2).

The existing kinetic equations for solitons derived in [Zakharov, 1971; El & Kamchatnov, 2005] account for solitons locations and their shifts in the collisions with other waves. However they disregard the soliton polarities, hence cannot distinguish the situations when the solitons reduce the field magnitude (when asynchronized) or sum up.

Focusing of an arbitrary number of solitons and breathers within the framework of the mKdV equation was considered in [Slunyaev & Pelinovsky, 2016] and is fully consistent with the reported in the present work conclusions. No details on the derivation were provided in the brief communication [Slunyaev & Pelinovsky, 2016]; it can be performed for the mKdV equation similar to the calculations described in the present paper, though significantly more straightforward. Hence, the large amplitude limit of the Gardner equation has been confirmed in [Slunyaev & Pelinovsky, 2016] directly. As the KdV equation possesses positive solitons only, they never produce large waves in the focal point (as shown in Fig. 1), according to (46). This conclusion is in qualitative agreement with the results of direct numerical simulations of irregular soliton ensembles [Shurgalina & Pelinovsky, 2016, 2017], which did not observe extreme wave events in the ensembles of unipolar solitons.

When dealing with the Darboux transform, the polarities of solitons and phases of breathers are specified by the seed functions. Though the expression for solution (16) may seem cumbersome, it is straightforward to be implemented numerically. For stability of the numerical solution it is useful to precalculate the differentiation of $\psi_j$ in the Wronskians analytically.

The considered scenarios of focusing of coherent structures are not limited by the framework of the long wave modes. We point out that the associated scattering problems for the mKdV equation and the focusing nonlinear Schrödinger equation (NLS) coincide when the solution of the NLS equation is a real function or possesses a constant complex phase [Wadati, 1973; Ablowitz et al, 1974]. Hence for any given instant $t$ the multisoliton solution of the mKdV equation $u(x,t)$ will possess the same spectral data of the scattering problem for the NLS equation. Moreover, an eigenvalue which describes a soliton in the mKdV equation will describe an envelope soliton of the NLS equation with the same amplitude. Pairs of complex conjugated discrete eigenvalues correspond to breathers of the mKdV equation, and to pairs of NLS solitons with opposite velocities and amplitudes twice smaller than the amplitude of the mKdV breather. Therefore, any multisoliton/multibreather solution of the mKdV equation in the focal point corresponds to a solution of the NLS equation, which is composed solely by envelope solitons with corresponding amplitudes. Thus the linear superposition of the partial soliton amplitudes in the focal point takes place in the NLS equation as well if the focused wave is cophased. (Though the evolution in time of the solutions of the mKdV and the NLS equations, of course, differs).

Furthermore, a similar correspondence occurs between the scattering problem for the mKdV equation for solutions on a constant pedestal and the problem for the focusing NLS with solitary perturbations of a condensate, if the NLS solution is uniphased (in particular, real-valued) at some time. Therefore any solution of the Gardner equation (11) at any instant $t$ corresponds to a multisoliton solution of the NLS equation on a condensate. (Envelope solitons of the NLS equation on a constant background are usually called breathers.) The scattering problem for solutions of the mKdV equation on a constant pedestal was transformed to the problem for the focusing Gardner equation on a zero background in [Grimshaw et al, 2010], therefore the problem of breathers of the NLS equation links straightforwardly to the results obtained in the present work. The relation between solitons of the NLS equation on a zero background and on a condensate, and the corresponding scattering data which correspond to these two cases were discussed in [Slunyaev, 2006]. It follows as a result that when solitons of the NLS equation propagating over a uniform condensate focus into a giant wave and get in phase with the condensate, the wave amplitude in the focal point is given by the linear superposition of amplitudes of the colliding solitons plus the amplitude of the condensate. This fact was shown in [Slunyaev et al, 2002; Slunyaev, 2006] for the case of one envelope soliton propagating over a condensate. The formula for the height of the $N$-order rational breather of the NLS equation, $2N + 1$, (suggested in [Akhmediev et al, 2009; Dubard & Matveev, 2013] and eventually proved in [Wang et al, 2017]), follows as a particular result of the described here link to the NLS framework. Obviously, such analogies can contribute to the studies of rogue-wave-type solutions of the NLS and mKdV equations [Ankiewicz & Akhmediev, 2018; Chen & Pelinovsky, 2018]. These issues will be discussed in more detail elsewhere.

Based on the discussed similarities of the solution in the focal point in a number of frameworks we suggest a hypothesis that the linear superposition of the partial amplitudes of soliton structures formulated in (46) takes place in a broader class of integrable equations.

## Acknowledgments

Support from RFBR Grant No. 18-02-00042 is acknowledged.

**Appendix A. Darboux transform for the focusing Gardner equation**

The method of construction of *N*-soliton solutions of the KdV equation with the help of the Darboux transform is well-known [Matveev & Salle, 1990]. It is based on the invariance of the Lax pair. In [Kulikov & Fraiman] this method was adapted for the modified KdV equation. In [Slunyaev & Pelinovsky, 1999] the method was applied to the Gardner equation with negative cubic nonlinearity (i.e., of defocusing type), and in [Slunyaev, 2001] – to the GE with positive cubic nonlinearity (1), though details of the modified scheme were not described. Here we provide the description of the method.

The KdV equation (5) is the compatibility condition for the so-called *L-A* Lax pair

$$\begin{cases} \hat{L}\psi = \lambda\psi \\ \hat{A}\psi = \partial_t \psi \end{cases}, \qquad (A.1)$$

$$\hat{L}\psi \equiv -\frac{\partial^2}{\partial x^2}\psi + v\psi, \quad \hat{A}\psi \equiv -4\frac{\partial^3}{\partial x^3}\psi + 6v\frac{\partial}{\partial x}\psi + 3\frac{\partial v}{\partial x}\psi.$$

The first equation in the system (A.1) is the stationary Schrödinger equation, which represents the associated linear scattering problem on eigenfunctions $\psi(x,t)$ and eigenvalues $\lambda$; the time $t$ is a parameter. The second equation in (A.1) describes the evolution in time of the functions $\psi(x,t)$. When eliminating the function $\psi$ from (A.1), one obtains the compatibility condition

$$\partial_t \hat{L} = \hat{A}\hat{L} - \hat{L}\hat{A}, \qquad (A.2)$$

which yields the KdV equation (5) for the function $v(x,t)$.

If, alternatively, the function $v(x,t)$ is eliminated from (A.1), then the compatibility condition for the Lax pair may be presented in the form of the modified KdV equation

$$w_t - 6\lambda w_x + 6w^2 w_x + w_{xxx} = 0 \qquad (A.3)$$

for the new function

$$w(x,t) = i\frac{\psi_x(x,t)}{\psi(x,t)}. \tag{A.4}$$

Bearing in mind the relation between the mKdV and the Gardner equations (3) it is straightforward to modify the change (A.4) to the following

$$u(x,t) = i\frac{\psi_x(x,t)}{\psi(x,t)} - \frac{1}{2}, \tag{A.5}$$

and then the compatibility condition for the Lax pair reads

$$u_t + \left(\frac{3}{2} - 6\lambda\right)u_x + 6uu_x + 6u^2 u_x + u_{xxx} = 0. \tag{A.6}$$

Equation (A.3) and (A.6) coincide with the mKdV equation (2) and the Gardner equation (1) respectively in appropriate moving references.

Thus, with the use of the changes (A.4) and (A.5) the solutions of the KdV equation may be transformed to the solutions of the mKdV and of the Gardner equations. Note that in the general case these relations may result in complex-valued solutions.

The generalized formulas for the *N*-fold application of the Darboux scheme read

$$\tilde{v}(x,t) = v(x,t) - 2\frac{\partial^2}{\partial x^2}\ln W_N(\psi_1, \psi_2, ..., \psi_N), \tag{A.7}$$

$$\tilde{\psi} = \frac{W_{N+1}(\psi_0, \psi_1, \psi_2, ..., \psi_N)}{W_N(\psi_1, \psi_2, ..., \psi_N)}. \tag{A.8}$$

In (A.7) the functions $v(x,t)$ and $\tilde{v}(x,t)$ are solutions of the KdV equation (A.6); $\psi_j(x,t)$, $j = 0, ..., N$, are particular solutions of the scattering problem (A.1) with the potential $v(x)$ for corresponding eigenvalues $\lambda_j$. The time dependence of functions $\psi_j(x,t)$ should meet the condition provided by the second equation in (A.1). Functions $\psi$ and $\tilde{\psi}$ are the solutions of the stationary Schrödinger equation for potentials $v(x)$ and $\tilde{v}(x)$ respectively. $W_N$ and $W_{N+1}$ denote the Wronskians given in (11). The combination of (A.5) and (A.8) results in the formula for the solutions of the Gardner equation (11)

$$u(x,t) = i\frac{\partial}{\partial x}\ln\frac{W_{N+1}}{W_N} - \frac{1}{2}. \tag{A.9}$$

For the seed functions (12) $\lambda_j = -\mu_j^2$ for $j = 1, ..., N$. In (13) we imply $\lambda_0 = 1/4$, and hence in (A.7) $v(x,t) = 0$, and equation (A.6) becomes identical to (1).

The multisoliton solution for the mKdV equation also follows immediately. If in (A.7) one selects $\lambda_0 = 0$ and $v(x,t) = 0$, then $\partial^2\psi_0/\partial x^2 = 0$, and the solution may be further simplified as suggested in [Kulikov & Fraiman, 1996]. In particular, the order of Wronskian $W_{N+1}$ may be reduced by one, and the solutions of the mKdV equation (2) are represented by

$$w(x,t) = -i\frac{\partial}{\partial x}\ln\frac{W_N(\psi'_1, \psi'_2, ..., \psi'_N)}{W_N(\psi_1, \psi_2, ..., \psi_N)}, \tag{A.10}$$

where primes mean the derivation with respect to *x*. This form of multisoliton solution was used in [Slunyaev & Pelinovsky, 2016] with somewhat different from (12) seed functions.

## Appendix B. Derivation of the local expression for the multisoliton solution

We assume that all the parameters $\mu_j$, $j = 1, ..., N$, are real and different. Let us consider the solution (11)-(13) assuming that the *n*-th soliton is located far from all the other solitons, having *L* solitons on the left and *R* solitons on the right (i.e., (23) holds), and the asymptotics (24) is valid. Then after simple manipulations the Wronskian $W_N$ reads

$$W_N = e^{\mu_1 x} \cdot \ldots \cdot e^{\mu_L x} e^{-\mu_r x} \cdot \ldots \cdot e^{-\mu_N x} \times$$

$$\times \begin{vmatrix} \mu_1^0 & \ldots & \mu_L^0 & \mu_n^0 \psi_n & (-\mu_r)^0 & \ldots & (-\mu_N)^0 \\ \mu_1^1 & \ldots & \mu_L^1 & \mu_n^1 \overline{\psi}_n & (-\mu_r)^1 & \ldots & (-\mu_N)^1 \\ \vdots & & \vdots & \vdots & \vdots & & \vdots \\ \mu_1^{N-1} & \ldots & \mu_L^{N-1} & \mu_n^{N-1} \cdot \begin{Bmatrix} \psi_n, & N \text{ is odd} \\ \overline{\psi}_n, & N \text{ is even} \end{Bmatrix} & (-\mu_r)^{N-1} & \ldots & (-\mu_N)^{N-1} \end{vmatrix}, \quad \text{(B.1)}$$

where $r \equiv N - R + 1$. With the use of identities

$$\psi = \frac{\psi + \overline{\psi}}{2} + \frac{\psi - \overline{\psi}}{2}, \qquad \overline{\psi} = \frac{\psi + \overline{\psi}}{2} - \frac{\psi - \overline{\psi}}{2} \qquad \text{(B.2)}$$

expression (B.1) may be transformed to the following,

$$W_N = \exp[\mu_1 x + \ldots + \mu_L x - \mu_r x - \ldots - \mu_N x] \left( D_1 \frac{\psi_n + \overline{\psi}_n}{2} + D_2 \frac{\psi_n - \overline{\psi}_n}{2} \right), \quad \text{(B.3)}$$

$$D_1 \equiv \begin{vmatrix} \mu_1^0 & \ldots & \mu_L^0 & \mu_n^0 & (-\mu_r)^0 & \ldots & (-\mu_N)^0 \\ \mu_1^1 & \ldots & \mu_L^1 & \mu_n^1 & (-\mu_r)^1 & \ldots & (-\mu_N)^1 \\ \vdots & & \vdots & \vdots & \vdots & & \vdots \\ \mu_1^{N-1} & \ldots & \mu_L^{N-1} & \mu_n^{N-1} & (-\mu_r)^{N-1} & \ldots & (-\mu_N)^{N-1} \end{vmatrix},$$

$$D_2 \equiv \begin{vmatrix} \mu_1^0 & \ldots & \mu_L^0 & (-\mu_n)^0 & (-\mu_r)^0 & \ldots & (-\mu_N)^0 \\ \mu_1^1 & \ldots & \mu_L^1 & (-\mu_n)^1 & (-\mu_r)^1 & \ldots & (-\mu_N)^1 \\ \vdots & & \vdots & \vdots & \vdots & & \vdots \\ \mu_1^{N-1} & \ldots & \mu_L^{N-1} & (-\mu_n)^{N-1} & (-\mu_r)^{N-1} & \ldots & (-\mu_N)^{N-1} \end{vmatrix}.$$

Then it is easy to calculate the derivative,

$$\partial_x W_N = [\mu_1 + \ldots + \mu_L - \mu_r - \ldots - \mu_N] W_n +$$
$$+ \mu_n \exp[\mu_1 x + \ldots + \mu_L x - \mu_r x - \ldots - \mu_N x] \left( D_1 \frac{\psi_n + \overline{\psi}_n}{2} - D_2 \frac{\psi_n - \overline{\psi}_n}{2} \right), \quad \text{(B.4)}$$

and hence

$$\frac{\partial_x W_N}{W_N} = [\mu_1 + \ldots + \mu_L - \mu_r - \ldots - \mu_N] + \mu_n \frac{\chi_n (\psi_n + \overline{\psi}_n) - (\psi_n - \overline{\psi}_n)}{\chi_n (\psi_n + \overline{\psi}_n) + (\psi_n - \overline{\psi}_n)}, \quad \text{(B.5)}$$

where

$$\chi_n \equiv \frac{D_1}{D_2}. \qquad \text{(B.6)}$$

The expression for $\chi_n$ may be easily guessed. In the general situation a determinant of a matrix should have the form of an algebraic polynomial of the same degree as the matrix size (see (17)). A determinant is zero when any two columns of the corresponding matrix are equal. Having the specific form of the matrices for $D_1, D_2$ (B.3) (see also the expressions for $M_\alpha$, $M_\beta$ (19), (21)), the sought polynomial is obviously the product of all combinations of differences between $d_j$ and $d_k, j \neq k$, which are in total $N - 1$ for each $d_j$:

$$\begin{vmatrix} d_1^0 & d_2^0 & \ldots & d_N^0 \\ d_1^1 & d_2^1 & \ldots & d_N^1 \\ \vdots & \vdots & \ddots & \vdots \\ d_1^{N-1} & d_2^{N-1} & \ldots & d_N^{N-1} \end{vmatrix} \propto (d_1 - d_2)(d_1 - d_3) \cdot \ldots \cdot (d_1 - d_N) \cdot (d_2 - d_3) \ldots (d_2 - d_N) \cdot \ldots \cdot (d_{N-1} - d_N). \quad \text{(B.7)}$$

The total number of roots of (B.7) is $N-1+N-2+\ldots+1 = N(N-1)/2$. The wanted polynomial gets reconstructed in (B.7) accurate to the common coefficient which has no significance. Applying (B.7) to $D_1$ and $D_2$ and making use of the fact that the expressions have a large common part, one may straightforwardly calculate (B.6) as

$$\chi_n = \frac{\prod_{j=1}^{L}(\mu_j - \mu_n) \prod_{j=N-R+1}^{N}(\mu_j + \mu_n)}{\prod_{j=1}^{L}(\mu_j + \mu_n) \prod_{j=N-R+1}^{N}(\mu_j - \mu_n)}, \qquad \chi_1 = 1 \text{ if } N = 1. \tag{B.8}$$

For different real-valued $\mu_j$ the number $\chi_n$ is real and differs from zero.

Following a similar way, we obtain

$$W_{N+1} = \exp[\mu_0 x + \mu_1 x + \ldots + \mu_L x - \mu_r x - \ldots - \mu_N x]\left(D_3 \frac{\psi_n + \overline{\psi}_n}{2} + D_4 \frac{\psi_n - \overline{\psi}_n}{2}\right), \tag{B.9}$$

$$D_3 \equiv \begin{vmatrix} \mu_0^0 & \mu_1^0 & \ldots & \mu_L^0 & \mu_n^0 & (-\mu_r)^0 & \ldots & (-\mu_N)^0 \\ \mu_0^1 & \mu_1^1 & \ldots & \mu_L^1 & \mu_n^1 & (-\mu_r)^1 & \ldots & (-\mu_N)^1 \\ \vdots & \vdots & & \vdots & \vdots & \vdots & & \vdots \\ \mu_0^N & \mu_1^N & \ldots & \mu_L^N & \mu_n^N & (-\mu_r)^N & \ldots & (-\mu_N)^N \end{vmatrix},$$

$$D_4 \equiv \begin{vmatrix} \mu_0^0 & \mu_1^0 & \ldots & \mu_L^0 & (-\mu_n)^0 & (-\mu_r)^0 & \ldots & (-\mu_N)^0 \\ \mu_0^1 & \mu_1^1 & \ldots & \mu_L^1 & (-\mu_n)^1 & (-\mu_r)^1 & \ldots & (-\mu_N)^1 \\ \vdots & \vdots & & \vdots & \vdots & \vdots & & \vdots \\ \mu_0^N & \mu_1^N & \ldots & \mu_L^N & (-\mu_n)^N & (-\mu_r)^N & \ldots & (-\mu_N)^N \end{vmatrix}$$

(we remind that $\mu_0 = -i/2$ and $r = N - R + 1$). Then

$$\frac{\partial_x W_{N+1}}{W_{N+1}} = [\mu_0 + \mu_1 + \ldots + \mu_L - \mu_r - \ldots - \mu_N] + \mu_n \frac{\chi_n e^{i\varphi_n}(\psi_n + \overline{\psi}_n) - (\psi_n - \overline{\psi}_n)}{\chi_n e^{i\varphi_n}(\psi_n + \overline{\psi}_n) + (\psi_n - \overline{\psi}_n)}, \tag{B.10}$$

where

$$e^{i\varphi_n} \equiv \frac{\mu_0 - \mu_n}{\mu_0 - \mu_n}. \tag{B.11}$$

The parameter $\varphi_n$ is real,

$$\varphi_n = \pi + \arctan \frac{4\mu_n}{4\mu_n^2 - 1}. \tag{B.12}$$

The solution (16) is calculated gathering (B.5) and (B.10), as

$$u(x,t) = i\mu_n \frac{s_n |\chi_n| e^{i\varphi_n}(\psi_n + \overline{\psi}_n) - (\psi_n - \overline{\psi}_n)}{s_n |\chi_n| e^{i\varphi_n}(\psi_n + \overline{\psi}_n) + (\psi_n - \overline{\psi}_n)} - i\mu_n \frac{s_n |\chi_n|(\psi_n + \overline{\psi}_n) - (\psi_n - \overline{\psi}_n)}{s_n |\chi_n|(\psi_n + \overline{\psi}_n) + (\psi_n - \overline{\psi}_n)}, \tag{B.13}$$

$$s_n \equiv \operatorname{sgn} \chi_n = \prod_{j=1, j \neq n}^{N}(\mu_j^2 - \mu_n^2), \qquad s_1 = 1 \text{ if } N = 1. \tag{B.14}$$

The relation (B.13) may be further simplified expanding the function $\psi_n$ (12), the eventual result depends on the sign of $\chi_n$ (B.14). In the case when the seed function is *cosh*, $\psi_n = \cosh(\Theta_n + i\theta_n)$, (B.13) yields

$$u = i\mu_n \cdot \begin{cases} \tanh\left(\Theta_n + i\left(\theta_n + \dfrac{\varphi_n}{2}\right) + \dfrac{1}{2}\ln\chi_n\right) - \tanh\left(\Theta_n + i\theta_n + \dfrac{1}{2}\ln\chi_n\right), & \text{if } \chi_n > 0 \\ \coth\left(\Theta_n + i\left(\theta_n + \dfrac{\varphi_n}{2}\right) + \dfrac{1}{2}\ln|\chi_n|\right) - \coth\left(\Theta_n + i\theta_n + \dfrac{1}{2}\ln|\chi_n|\right), & \text{if } \chi_n < 0 \end{cases}. \quad (B.15)$$

When $\psi_n = \sinh(\Theta_n + i\theta_n)$, then in (B.15) functions *tanh* and *coth* should be swapped.

The solution $u(x,t)$ (B.15) is real-valued for arbitrary $\mu_n$, if $\theta_n + \varphi_n/2 = -\theta_n$, what imposes the following condition on $\theta_n$, obtained after some trigonometric transformations,

$$2i\mu_n = \tanh 2i\theta_n, \quad (B.16)$$

Which is equivalent to (15). As a result, (B.15) in the situation $\psi_n = \cosh(\Theta_n + i\theta_n))$ gives the following,

$$u(x,t) = \begin{cases} i\mu_n \tanh\left(\Theta_n - i\theta_n + \dfrac{1}{2}\ln\chi_n\right) + c.c., & \text{if } \chi_n > 0 \\ i\mu_n \coth\left(\Theta_n - i\theta_n + \dfrac{1}{2}\ln|\chi_n|\right) + c.c., & \text{if } \chi_n < 0 \end{cases}. \quad (B.17)$$

When $\psi_n$ is a *sinh* function, functions *tanh* and *coth* in (B.17) should be swapped.

If the seed functions are presented in the form (14), the solution (B.17) may be written in the following general form,

$$u(x,t) = i\mu_n \tanh\left(\Theta_n - i\theta_n - i\dfrac{\pi}{2}\left(l_n + \dfrac{1-s_n}{2}\right) + \dfrac{1}{2}\ln|\chi_n|\right) + c.c. \quad (B.18)$$

## Appendix C. Derivation of the multisoliton / multibreather solution in the focal point

Let us calculate the solution in the focal point $u(x = 0, t = 0)$ directly from (16), assuming that parameters $\mu_j$, $j = 1, \ldots, N$, are different but may be complex numbers.

From (18) one straightforwardly obtains

$$\partial_x W_N = \sum_\alpha \mu_{\alpha_2}\mu_{\alpha_3}^2\mu_{\alpha_4}^3 \cdot \ldots \cdot \mu_{\alpha_N}^{N-1} \cdot \mu_{\alpha_1}\overline{\psi}_{\alpha_1}\overline{\psi}_{\alpha_2}\psi_{\alpha_3}\overline{\psi}_{\alpha_4}\cdot\ldots\cdot\begin{cases}\psi_{\alpha_N}, & N \text{ is odd} \\ \overline{\psi}_{\alpha_N}, & N \text{ is even}\end{cases} +$$

$$+\sum_\alpha \mu_{\alpha_2}\mu_{\alpha_3}^2\mu_{\alpha_4}^3 \cdot \ldots \cdot \mu_{\alpha_N}^{N-1} \cdot \psi_{\alpha_1}\mu_{\alpha_2}\psi_{\alpha_2}\psi_{\alpha_3}\overline{\psi}_{\alpha_4}\cdot\ldots\cdot\begin{cases}\psi_{\alpha_N}, & N \text{ is odd} \\ \overline{\psi}_{\alpha_N}, & N \text{ is even}\end{cases} + \ldots$$

$$\ldots + \sum_\alpha \mu_{\alpha_2}\mu_{\alpha_3}^2\mu_{\alpha_4}^3 \cdot \ldots \cdot \mu_{\alpha_N}^{N-1} \cdot \psi_{\alpha_1}\overline{\psi}_{\alpha_2}\psi_{\alpha_3}\overline{\psi}_{\alpha_4}\cdot\ldots\cdot \mu_{\alpha_N}\begin{cases}\overline{\psi}_{\alpha_N}, & N \text{ is odd} \\ \psi_{\alpha_N}, & N \text{ is even}\end{cases}. \quad (C.1)$$

It is easy to see that the first summand cancels due to the permutation $(\alpha_1, \alpha_2) \to (\alpha_2, \alpha_1)$. Similarly, the second summand cancels due to the permutation $(\alpha_2, \alpha_3) \to (\alpha_3, \alpha_2)$, and so on. As a result, only the last summand in the expression (C.1) retains,

$$\partial_x W_N = \sum_\alpha c_\alpha M_\alpha \Psi_\alpha(x,t) \cdot \mu_{\alpha_N}\begin{cases}\overline{\psi}_{\alpha_N}/\psi_{\alpha_N}, & N \text{ is odd} \\ \psi_{\alpha_N}/\overline{\psi}_{\alpha_N}, & N \text{ is even}\end{cases} \quad (C.2)$$

((C.2) is valid for any $x$ and $t$).

Consider the product

$$G \equiv \left(\sum_\alpha c_\alpha M_\alpha \Psi_\alpha(0,0)\right) \cdot \left(\sum_{j=1}^N \mu_j \begin{cases}\overline{\psi}_j(i\theta_j)/\psi_j(i\theta_j), & N \text{ is odd} \\ \psi_j(i\theta_j)/\overline{\psi}_j(i\theta_j), & N \text{ is even}\end{cases}\right), \quad (C.3)$$

putting $x = 0$, $t = 0$. The first summation in (C.3) is performed for all permutations of indices $\alpha$, though the second multiplier represents an ordinary summation. It may be realized that the product $G$ is identical to the following expression, which consists of summations through permutations of $\alpha$,

$$G = \sum_\alpha \mu_{\alpha_2} \mu_{\alpha_3}^2 \cdot \ldots \cdot \mu_{\alpha_N}^{N-1} \cdot \overline{\psi}_{\alpha_2} \psi_{\alpha_3} \overline{\psi}_{\alpha_4} \cdot \ldots \cdot \mu_{\alpha_1} \begin{cases} \overline{\psi}_{\alpha_1} \psi_{\alpha_N}, & N \text{ is odd} \\ \dfrac{\psi_{\alpha_1}^2}{\overline{\psi}_{\alpha_1}} \overline{\psi}_{\alpha_N}, & N \text{ is even} \end{cases} +$$

$$+ \sum_\alpha \mu_{\alpha_2} \mu_{\alpha_3}^2 \mu_{\alpha_4}^3 \cdot \ldots \cdot \mu_{\alpha_N}^{N-1} \cdot \psi_{\alpha_1} \psi_{\alpha_3} \overline{\psi}_{\alpha_4} \cdot \ldots \cdot \mu_{\alpha_2} \begin{cases} \dfrac{\overline{\psi}_{\alpha_2}^2}{\psi_{\alpha_2}} \psi_{\alpha_N}, & N \text{ is odd} \\ \psi_{\alpha_2} \overline{\psi}_{\alpha_N}, & N \text{ is even} \end{cases} + \ldots$$

$$\ldots + \sum_\alpha \mu_{\alpha_2} \mu_{\alpha_3}^2 \cdot \ldots \cdot \mu_{\alpha_N}^{N-1} \cdot \psi_{\alpha_1} \overline{\psi}_{\alpha_2} \psi_{\alpha_3} \cdot \ldots \cdot \mu_{\alpha_N} \begin{cases} \overline{\psi}_{\alpha_N}, & N \text{ is odd} \\ \psi_{\alpha_N}, & N \text{ is even} \end{cases} \quad \text{at} \quad (x = 0, t = 0). \quad \text{(C.4)}$$

To simplify the expression (C.4) further, we rewrite the equality (8) with the use of (7) and (29) as

$$i\mu_j \left( \frac{\psi_j(i\theta_j)}{\overline{\psi}_j(i\theta_j)} + \frac{\overline{\psi}_j(i\theta_j)}{\psi_j(i\theta_j)} \right) = 1, \quad j = 0, \ldots, N. \quad \text{(C.5)}$$

Then (C.4) transforms to

$$G = \sum_\alpha \mu_{\alpha_2} \mu_{\alpha_3}^2 \cdot \ldots \cdot \mu_{\alpha_N}^{N-1} \cdot \overline{\psi}_{\alpha_2} \psi_{\alpha_3} \overline{\psi}_{\alpha_4} \cdot \ldots \cdot \begin{cases} \mu_{\alpha_1} \overline{\psi}_{\alpha_1} \psi_{\alpha_N}, & N \text{ is odd} \\ (-i - \mu_{\alpha_1} \overline{\psi}_{\alpha_1}) \overline{\psi}_{\alpha_N}, & N \text{ is even} \end{cases} +$$

$$+ \sum_\alpha \mu_{\alpha_2} \mu_{\alpha_3}^2 \mu_{\alpha_4}^3 \cdot \ldots \cdot \mu_{\alpha_N}^{N-1} \cdot \psi_{\alpha_1} \psi_{\alpha_3} \overline{\psi}_{\alpha_4} \cdot \ldots \cdot \begin{cases} (-i - \mu_{\alpha_2} \psi_{\alpha_2}) \psi_{\alpha_N}, & N \text{ is odd} \\ \mu_{\alpha_2} \psi_{\alpha_2} \overline{\psi}_{\alpha_N}, & N \text{ is even} \end{cases} + \ldots$$

$$\ldots + \sum_\alpha \mu_{\alpha_2} \mu_{\alpha_3}^2 \cdot \ldots \cdot \mu_{\alpha_N}^{N-1} \cdot \psi_{\alpha_1} \overline{\psi}_{\alpha_2} \psi_{\alpha_3} \cdot \ldots \cdot \mu_{\alpha_N} \begin{cases} \overline{\psi}_{\alpha_N}, & N \text{ is odd} \\ \psi_{\alpha_N}, & N \text{ is even} \end{cases} \quad \text{at} \quad (x = 0, t = 0). \quad \text{(C.6)}$$

Due to the permutations $(\alpha_1, \alpha_2) \to (\alpha_2, \alpha_1)$, $(\alpha_2, \alpha_3) \to (\alpha_3, \alpha_2)$ and so on, only the summands with imaginary units retain in (C.6) in addition to the last line of (C.6), which gives exactly $\partial_x W_N$ (see (C.1) and (C.2)). Therefore (C.6) yields the following relation

$$\frac{\partial_x W_N}{W_N} = i \cdot m + \sum_{j=1}^N \mu_j \begin{cases} \overline{\psi}_j / \psi_j, & N \text{ is odd} \\ \psi_j / \overline{\psi}_j, & N \text{ is even} \end{cases} \quad \text{at} \quad (x = 0, t = 0), \quad m = \left[ \frac{N}{2} \right], \quad \text{(C.7)}$$

where the number $m$ is the integer part of the ratio $N/2$.

Note that the equality (C.5) is also valid for $j = 0$, therefore the remaining part of the solution (16) follows straightforwardly similar to (C.7) with the use of the definition $\mu_0 = -i/2$:

$$\frac{\partial_x W_{N+1}}{W_{N+1}} = -\frac{i}{2} + i \cdot p + \sum_{j=1}^N \mu_j \begin{cases} \psi_j / \overline{\psi}_j, & N \text{ is odd} \\ \overline{\psi}_j / \psi_j, & N \text{ is even} \end{cases} \quad \text{at} \quad (x = 0, t = 0), \quad p = \left[ \frac{N+1}{2} \right]. \quad \text{(C.8)}$$

Combining (C.7) and (C.8), we obtain the solution

$$u = \begin{cases} -1 + i\sum_{j=1}^{N} \mu_j \left( \psi_j / \overline{\psi}_j - \overline{\psi}_j / \psi_j \right), & N \text{ is odd} \\ i\sum_{j=1}^{N} \mu_j \left( \overline{\psi}_j / \psi_j - \psi_j / \overline{\psi}_j \right), & N \text{ is even} \end{cases} \quad \text{at} \quad (x = 0, t = 0). \quad (C.9)$$

With the use of (C.5) the solution (C.9) may be further reduced to the form,

$$u = \begin{cases} -2i\mu_1 \dfrac{\overline{\psi}_1}{\psi_1} + 2i\mu_2 \dfrac{\psi_2}{\overline{\psi}_2} - 2i\mu_3 \dfrac{\overline{\psi}_3}{\psi_3} + \ldots - 2i\mu_N \dfrac{\overline{\psi}_N}{\psi_N}, & N \text{ is odd} \\ -2i\mu_1 \dfrac{\psi_1}{\overline{\psi}_1} + 2i\mu_2 \dfrac{\overline{\psi}_2}{\psi_2} - 2i\mu_3 \dfrac{\psi_3}{\overline{\psi}_3} + \ldots + 2i\mu_N \dfrac{\overline{\psi}_N}{\psi_N} & N \text{ is even} \end{cases} \quad \text{at} \quad (x = 0, t = 0), \quad (C.10)$$

which is replicated in (32).

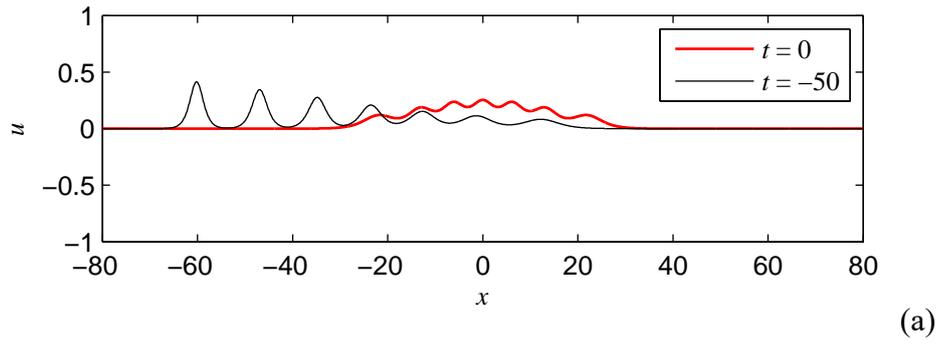

(a)

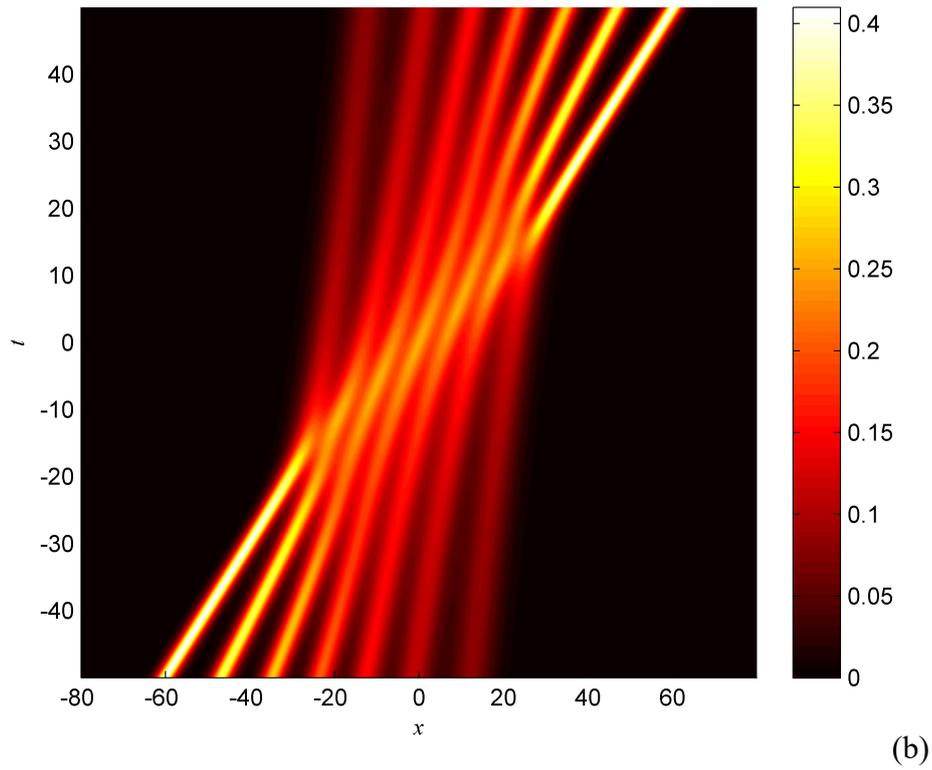

(b)

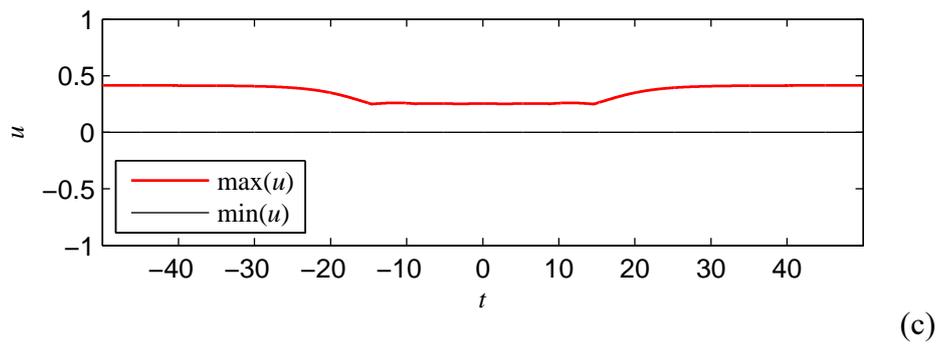

(c)

Fig. 1. Interaction of 7 positive solitons of the Gardner equation: the instants long before focusing ($t = -50$) and at the moment when focusing should occur ($t = 0$) (a); the spatiotemporal plot of $u(x,t)$ (b); and the corresponding temporal evolution of maximum and minimum (c). The soliton parameters are $2\mu = \{1, 0.9, 0.8, 0.7, 0.6, 0.5, 0.4\}$.

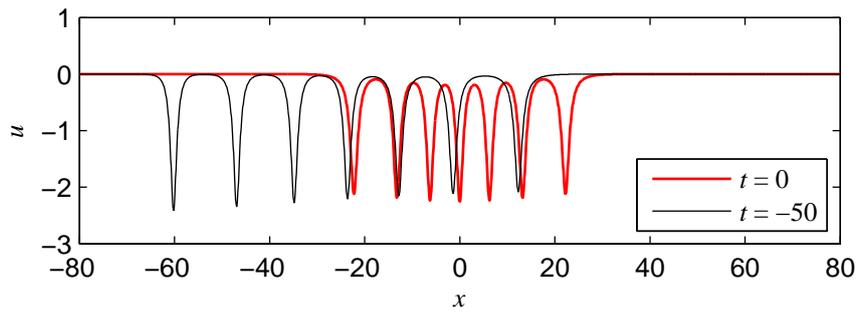

(a)

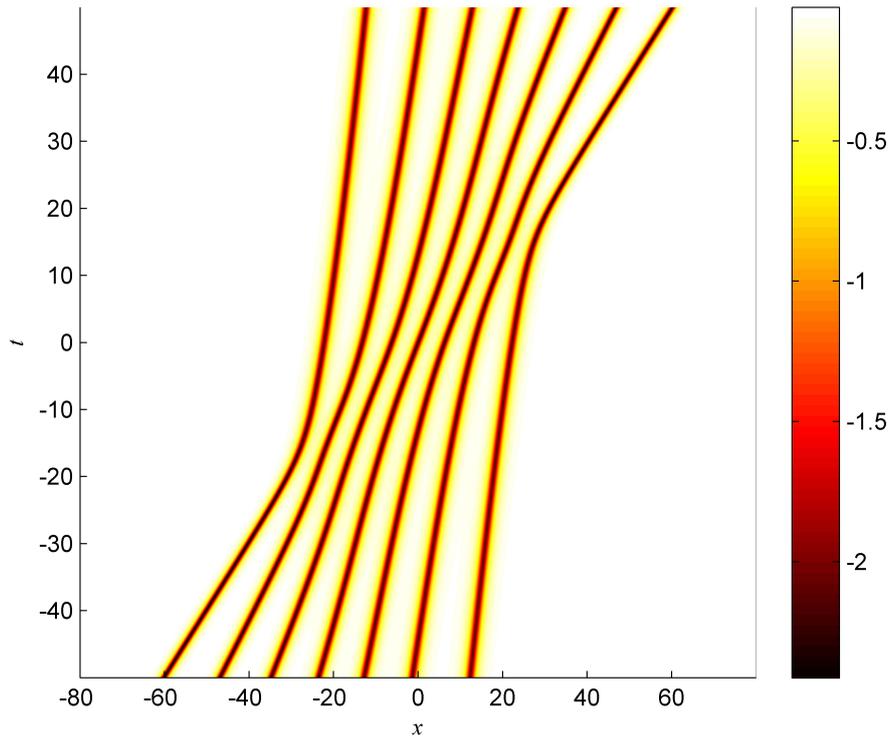

(b)

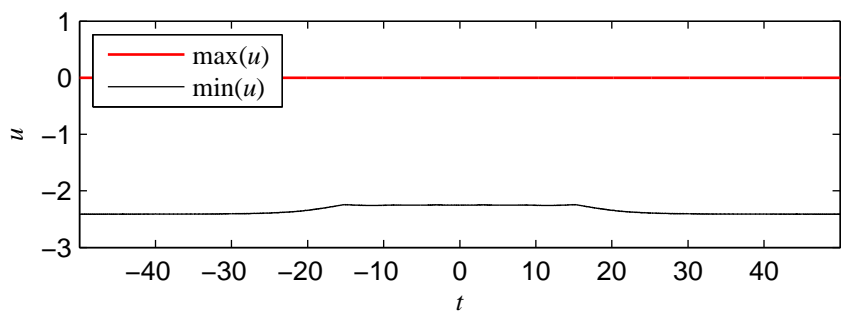

(c)

Fig. 2. Interaction of 7 negative solitons with the same parameters $\mu$ as in Fig. 1.

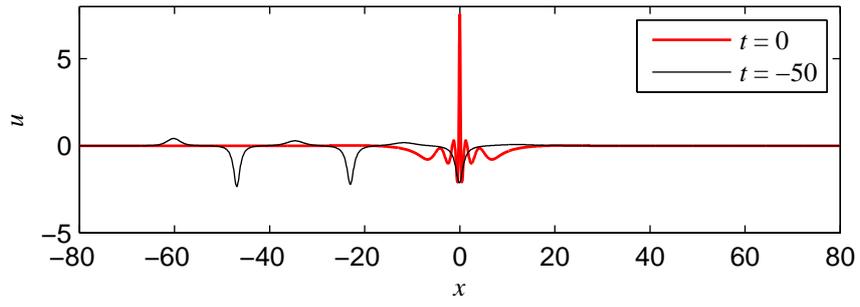

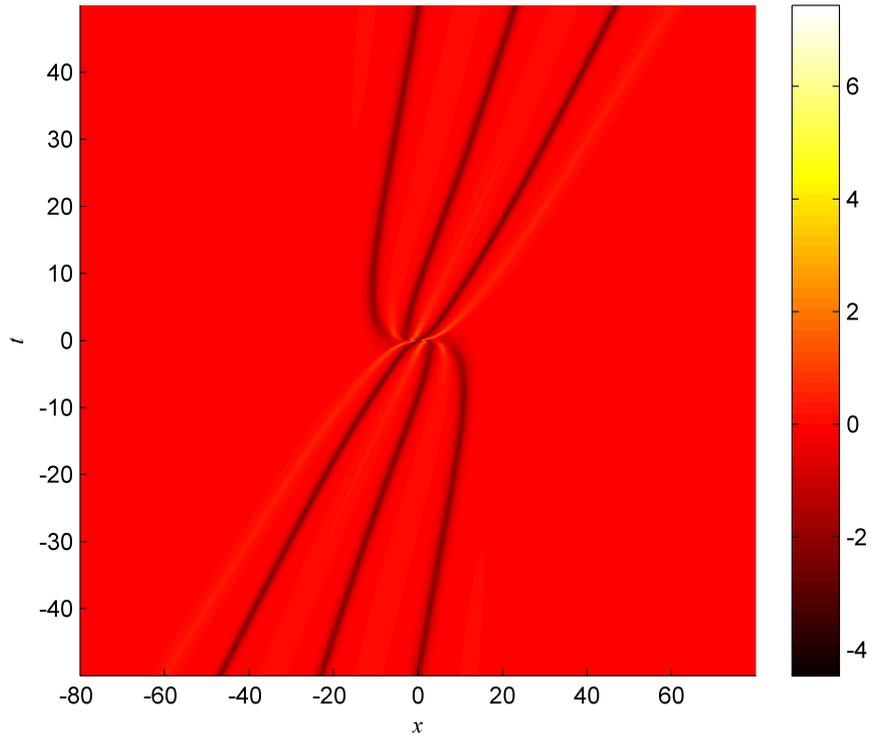

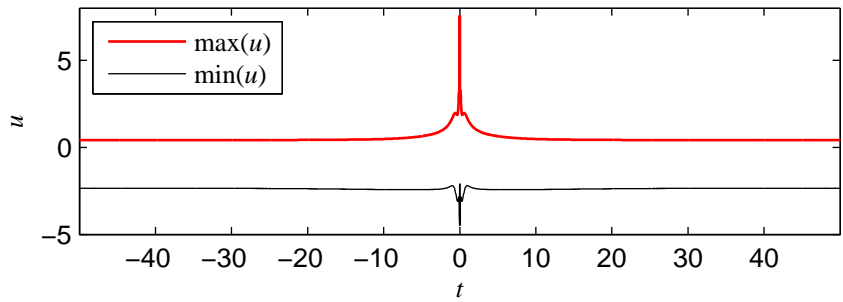

Fig. 3. Interaction of 7 solitons of alternating polarities with the same parameters $\mu$ as in Figs. 1,2. The fastest soliton is positive.

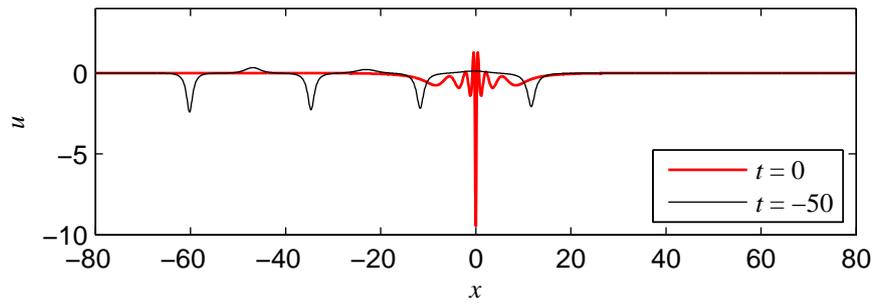

(a)

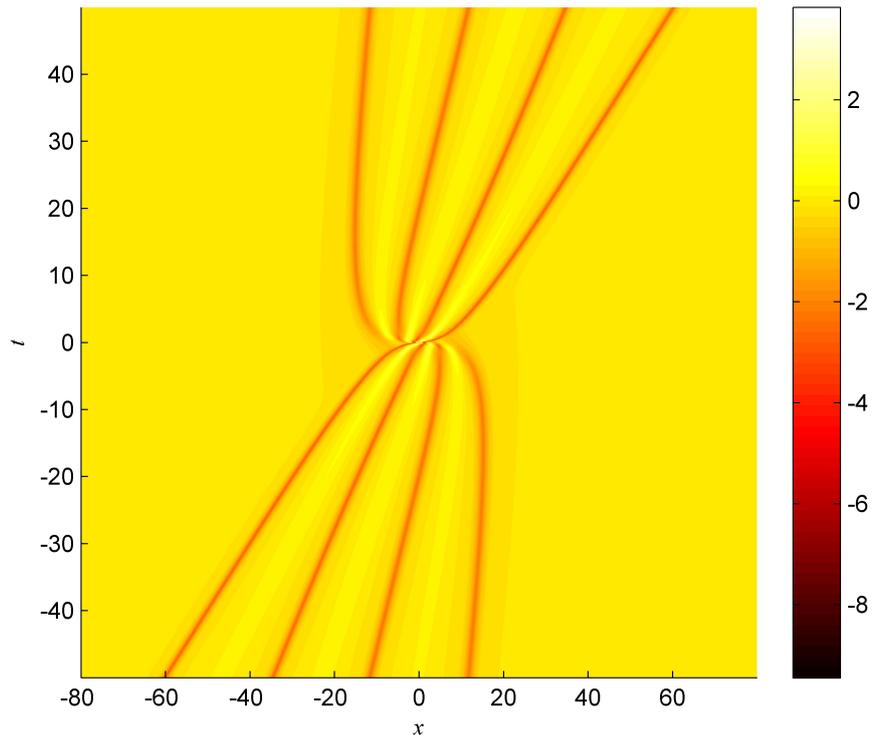

(b)

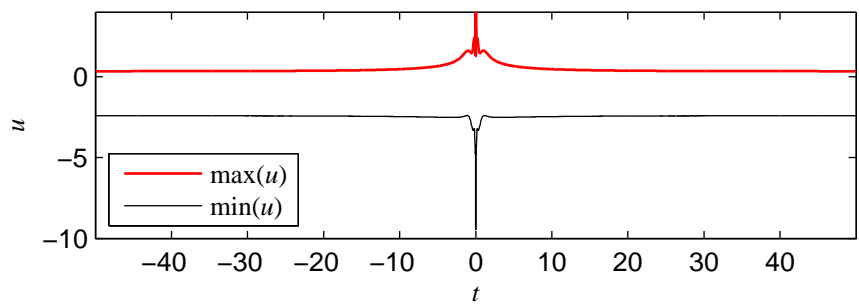

(c)

Fig. 4. Same as in Fig. 3, but the solitons have opposite signs (hence the fastest soliton is negative).

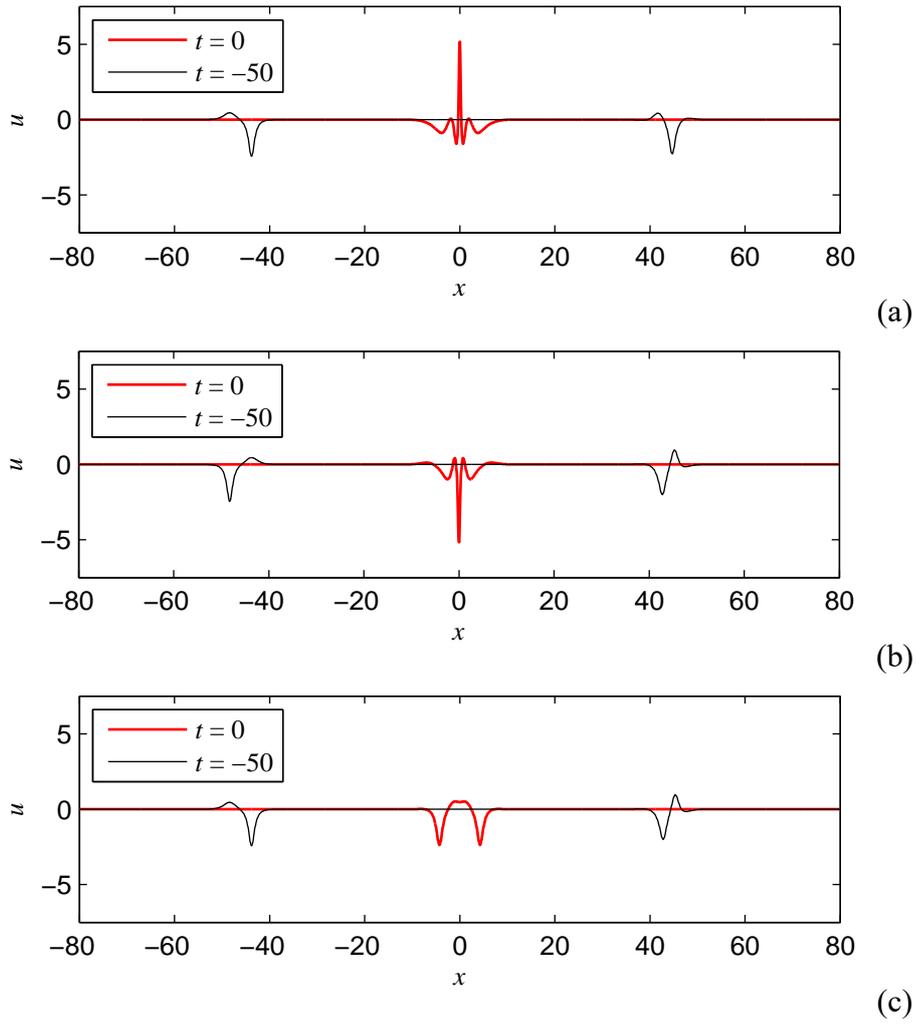

Fig. 5. Focusing of two breathers of the Gardner equation with parameters $2\mu = 1 \pm 0.2i$, $|B_1| \approx 2.8$ (from the left at $t = -50$) and $2\mu = 0.8 \pm 0.7i$, $|B_2| \approx 2.35$ (from the right at $t = -50$). Different panels correspond to different phases of the breathers, which provide different signs of $B_1$ and $B_2$: $B_1 > 0$, $B_2 > 0$ (a), $B_1 < 0$, $B_2 < 0$ (b), $B_1 > 0$, $B_2 < 0$ (c).

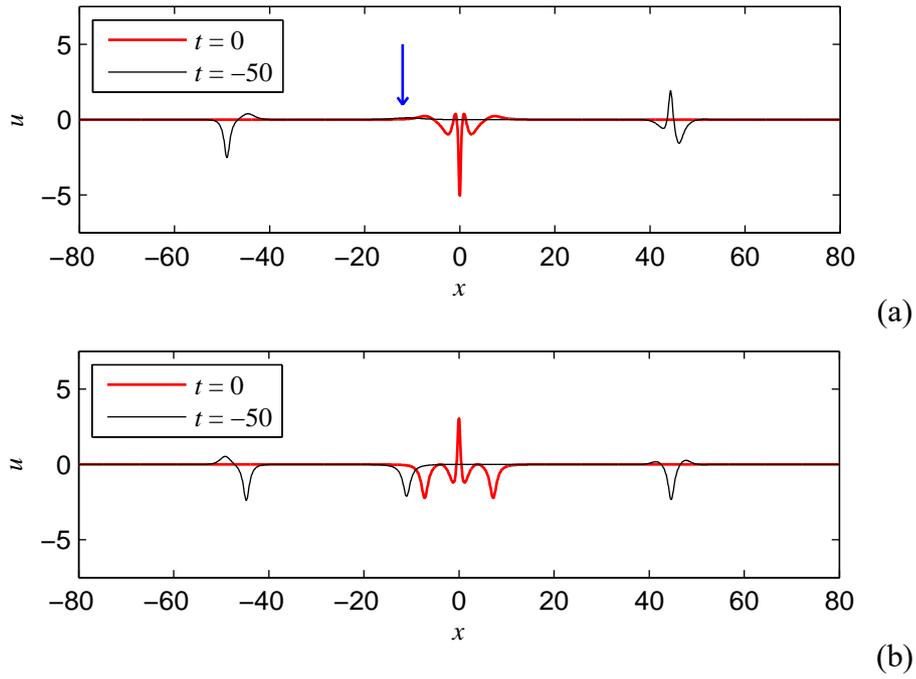

Fig. 6. Focusing of two breathers of the Gardner equation (with the same parameters as in Fig. 5) and one soliton with parameter $2\mu = 0.5$. (a): breathers with $B_1 > 0$, $B_2 > 0$ and one positive soliton, $A_1 \approx 0.12$ (shown with the arrow); (b): breathers with $B_1 < 0$, $B_2 < 0$ and one negative soliton, $A_1 \approx -2.12$.

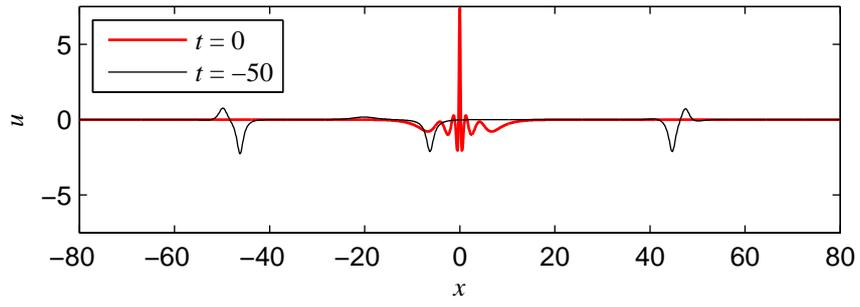

(a)

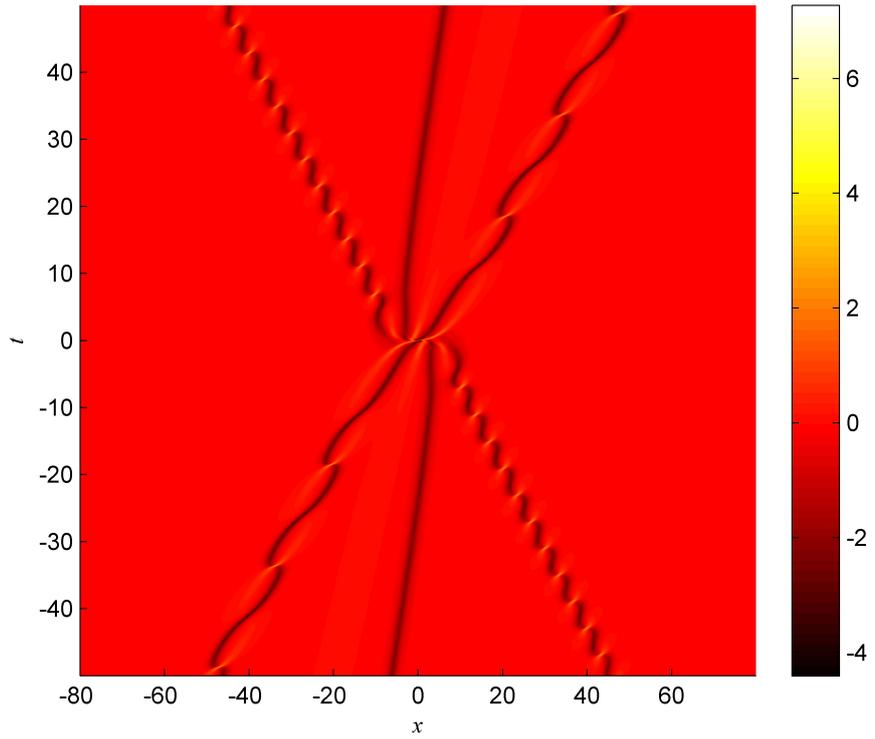

(b)

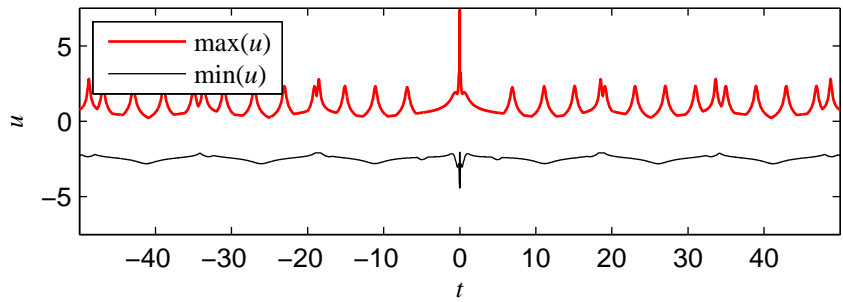

(c)

Fig. 7. Optimal focusing of two breathers $2\mu = 1 \pm 0.2i$ ($B_1 \approx 2.8$) and $2\mu = 0.8 \pm 0.7i$ ($B_2 \approx 2.35$) and two solitons $2\mu = 0.6$ ($A_1 \approx 0.17$) and $2\mu = 0.5$ ($A_2 \approx -2.12$): the instants long before focusing and at the focusing (a); the spatiotemporal plot of $u(x,t)$ (b), and the corresponding temporal evolution of maximum and minimum (c).

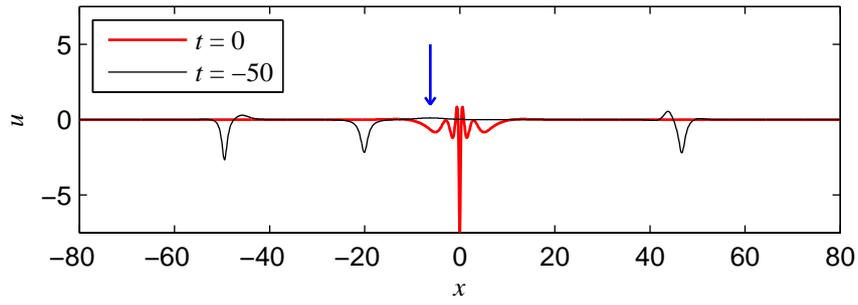

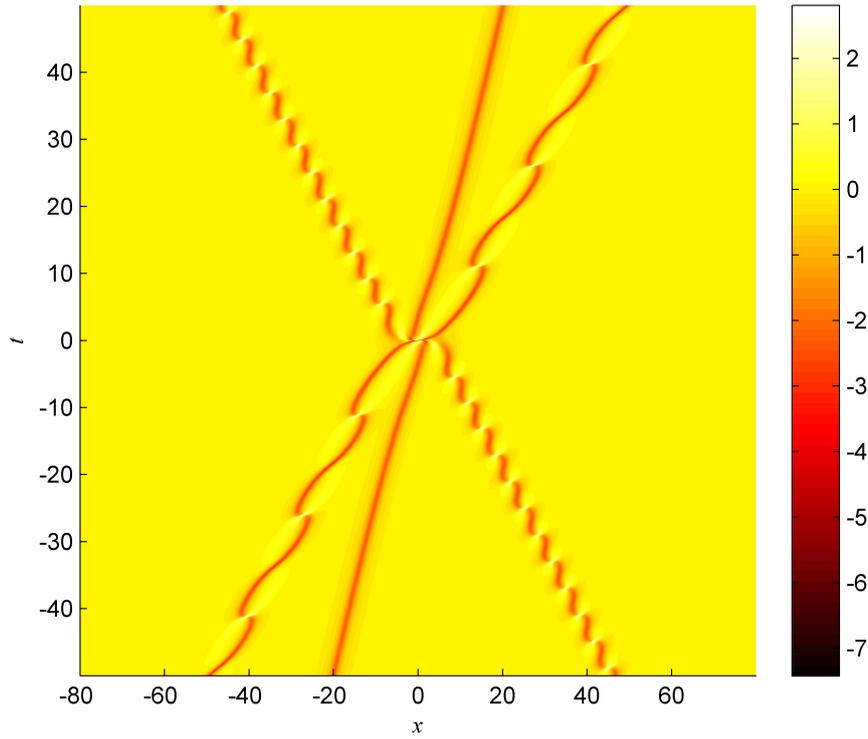

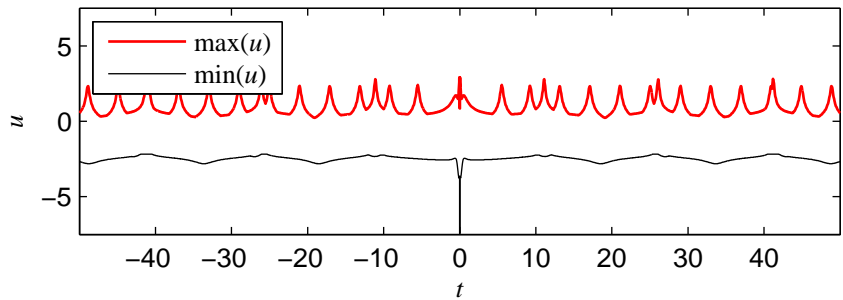

Fig. 8. Optimal focusing of two breathers and two solitons with the same parameters $\mu$ as in Fig. 7 but all the polarities and phases are opposite to that case. The solitons have amplitudes $A_1 \approx -2.17$ and $A_2 \approx 0.12$ (the latter is shown with the arrow in panel (a)), the breathers are characterized by values $B_1 \approx -2.8$ and $B_2 \approx -2.35$.